\documentclass[twocolumn]{aastex62}

\usepackage{amsmath}
\usepackage{subfigure}
\usepackage{amssymb}
\usepackage{url}

\shorttitle{DTD of RR Lyrae}
\shortauthors{Sarbadhicary et al.}

\begin{document}
\graphicspath{{./Sarba18_RRLyraeDTD_Plots/}}

\title{The RR Lyrae Delay-Time Distribution: A Novel Perspective on Models of Old Stellar Populations}

\correspondingauthor{Sumit K. Sarbadhicary}
\email{sarbadhi@msu.edu}

\author[0000-0002-4781-7291]{Sumit K. Sarbadhicary}
\affiliation{Department of Physics and Astronomy, Michigan State University, East Lansing, MI 48824, USA}

\author{Mairead Heiger}
\affiliation{Pittsburgh Particle Physics, Astrophysics and Cosmology Center (PITT PACC), University of Pittsburgh, 3941
O'Hara St, Pittsburgh, PA 15260, USA}

\author[0000-0003-3494-343X]{Carles Badenes}
\affiliation{Pittsburgh Particle Physics, Astrophysics and Cosmology Center (PITT PACC), University of Pittsburgh, 3941
O'Hara St, Pittsburgh, PA 15260, USA}

\author[0000-0002-6330-2394]{Cecilia~Mateu}
\affiliation{Departamento de Astronom\'ia, Facultad de Ciencias, Universidad de la Rep\'ublica, Igu\'a 4225, 14000, Montevideo, Uruguay}

\author{Jeffrey Newman}
\affiliation{Pittsburgh Particle Physics, Astrophysics and Cosmology Center (PITT PACC), University of Pittsburgh, 3941
O'Hara St, Pittsburgh, PA 15260, USA}

\author{Robin Ciardullo}
\affiliation{Department of Astronomy \& Astrophysics, The Pennsylvania State University, University Park, PA 16802, USA}

\author[0000-0002-0430-7793]{Na'ama Hallakoun}
\affiliation{Department of Particle Physics and Astrophysics, Weizmann Institute of Science, Rehovot, 7610001, Israel}

\author{Dan Maoz}
\affiliation{School of Physics and Astronomy, Tel-Aviv University, Tel-Aviv 6997801, Israel}

\author[0000-0002-8400-3705]{Laura Chomiuk}
\affiliation{Department of Physics and Astronomy, Michigan State University, East Lansing, MI 48824, USA}


\begin{abstract}
The delay-time distribution (DTD) is the occurrence rate of a class of objects as a function of time after a hypothetical burst of star formation. DTDs are mainly used as a statistical test of stellar evolution scenarios for supernova progenitors, but they can be applied to many other classes of astronomical objects. We calculate the first
DTD for RR Lyrae variables using 29,810 RR Lyrae from the OGLE-IV survey and  a map of the stellar-age distribution (SAD) in the Large Magellanic Cloud (LMC).  We find that $\sim 46\%$ of the OGLE-IV RR Lyrae are associated with delay-times older than 8 Gyr (main-sequence progenitor masses less than 1 M$_{\odot}$), and consistent with existing constraints on their ages, but surprisingly about $51\%$ of RR Lyrae appear have delay times $1.2-8$ Gyr (main-sequence masses between $1 - 2$ M$_{\odot}$ at LMC metallicity). This intermediate-age signal also persists outside the Bar-region where crowding is less of a concern, and we verified that without this signal, the spatial distribution of the OGLE-IV RR Lyrae is inconsistent with the SAD map of the LMC. Since an intermediate-age RR Lyrae channel is in tension with the lack of RR Lyrae in intermediate-age clusters (noting issues with small-number statistics), and the age-metallicity constraints of LMC stars, our DTD result possibly indicates that systematic uncertainties may still exist in SAD measurements of old-stellar populations, perhaps stemming from the construction methodology or the stellar evolution models used. We described tests to further investigate this issue.

\end{abstract}

\section{Introduction}

A detailed understanding of stellar evolution remains one of the most sought-after goals in astrophysics. Popular stellar
evolution codes such as Geneva \citep{Schaller1992, Schaerer1993}, Y$^{\rm{2}}$ \citep{Kim2002, Yi2003, Demarque2004},
BaSTI \citep{Pietrinferni2004, Pietrinferni2006, Hidalgo2018}, Darthmouth \citep{Dotter2008}, PARSEC \citep{Bressan2012, Chen2014}
and MESA \citep{Paxton2011, Paxton2013, Paxton2015, Paxton2018} are powerful tools for interpreting observations of
stellar populations. However, many essential topics in stellar evolution are still not well understood, and/or not properly taken into account in even the most state-of-the-art models. Examples of such topics include convection, mass loss and mass transfer, common envelope
evolution, and binary interaction. Often, these three-dimensional phenomena are approximated by 
simplified parametric models tuned to specific observables and integrated into one-dimensional stellar evolution codes. These uncertainties
limit our understanding of many important phases of stellar evolution, such as the horizontal branch, the asymptotic giant branch (AGB) and post-AGB phase, planetary nebulae, and supernovae \citep[see discussions in][]{Gallart2005, Conroy2009, Conroy2013}.

The delay-time distribution (DTD) is a promising method for testing stellar evolution models in complex stellar populations
\citep{MaozMan2012, Maoz2014}. The DTD is defined as the occurrence rate of a class of astronomical object as a function of time
since a hypothetical brief burst of star formation; it is equivalent to the impulse response, or Green's function. The DTD
constrains the evolutionary timescale and formation efficiency of the object's progenitors, and theoretical DTDs are common
predictions of stellar population synthesis models \citep{Mennekens2010, Nelemans2013, Toonen2013, Zapartas17}. Observationally,
DTDs can be derived from surveys of objects, provided that the stellar age distributions (SADs) of their host galaxies are measured \citep{GalYam2004, Totani2008, Maoz2010a, Maoz2011, Maoz2012b, Graur2014, Maoz2017, Friedmann2018}. 

More recently it was shown that DTDs can be a useful stellar evolution diagnostic in Local Group galaxies with high-quality observations of their resolved stellar populations \citep{Badenes2010, Maoz2010, Badenes15}. Using this approach, \cite{Badenes15} measured the first
DTD for planetary nebula, showing evidence of two distinct populations of planetary nebula progenitors: one with ages of 5--8 Gyrs, and another with ages of 35--800 Myrs. The key advantage of a DTD is that it constrains the evolutionary timescales for the progenitors of the entire population of
objects in a galaxy, taking into account the variety of star-formation histories that these objects have evolved from. It can be a powerful tool for identifying or ruling out the presence of specific formation channels, measuring their efficiency, and identifying physical mechanisms that are not part of canonical progenitor models.

\begin{figure*} 
	\centering
	\includegraphics[width=0.7\textwidth]{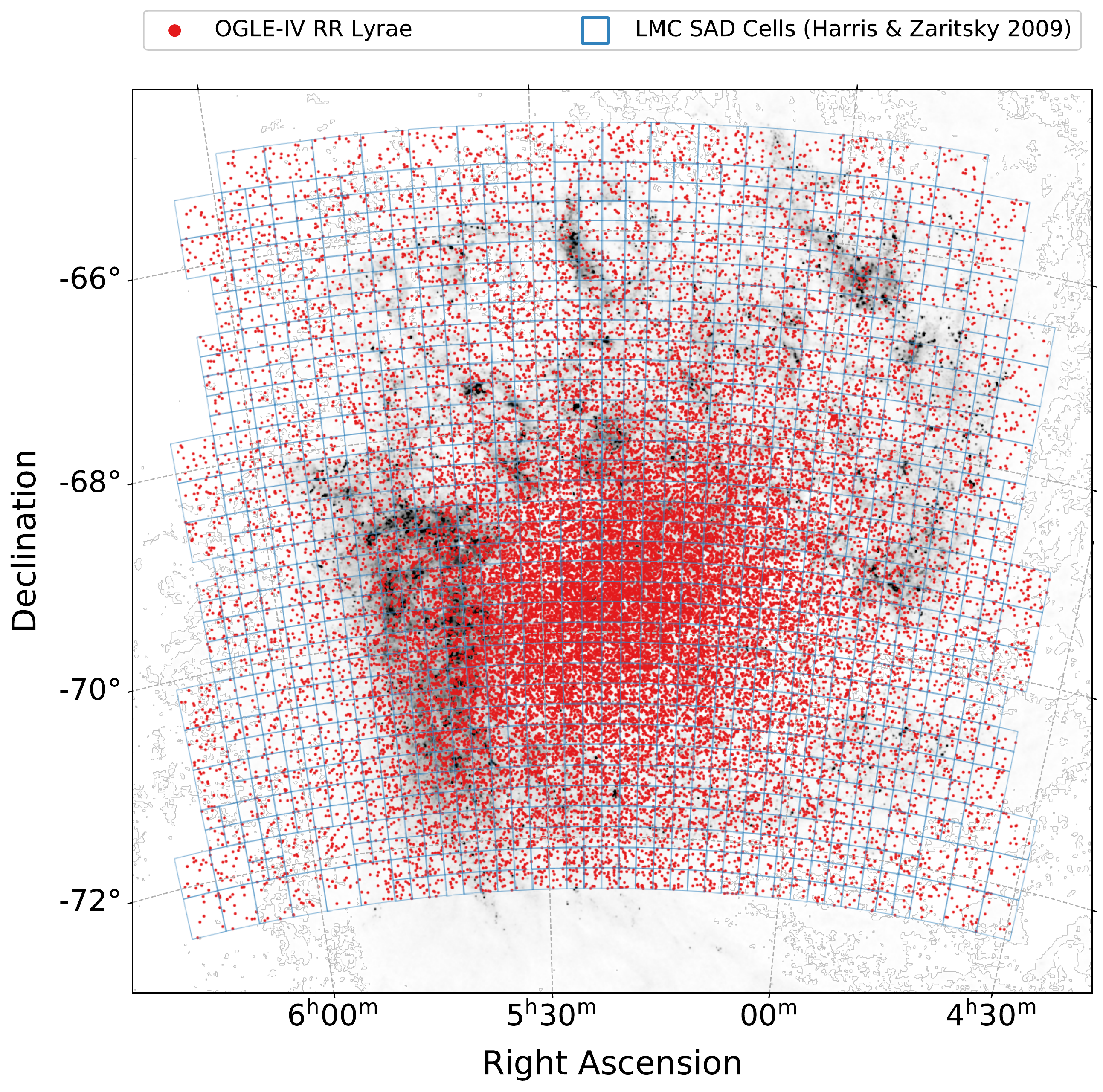}
	\caption{The spatial distribution of LMC RR Lyrae (red dots) in the OGLE-IV sample overlaid on the spatial cells (blue) from the \cite{Harris2009} SAD map of the galaxy.}
	\label{fig:rrlmap}
      \end{figure*}
      
\begin{figure}
    \hspace{-0.3in}
	\includegraphics[width=1.2\columnwidth]{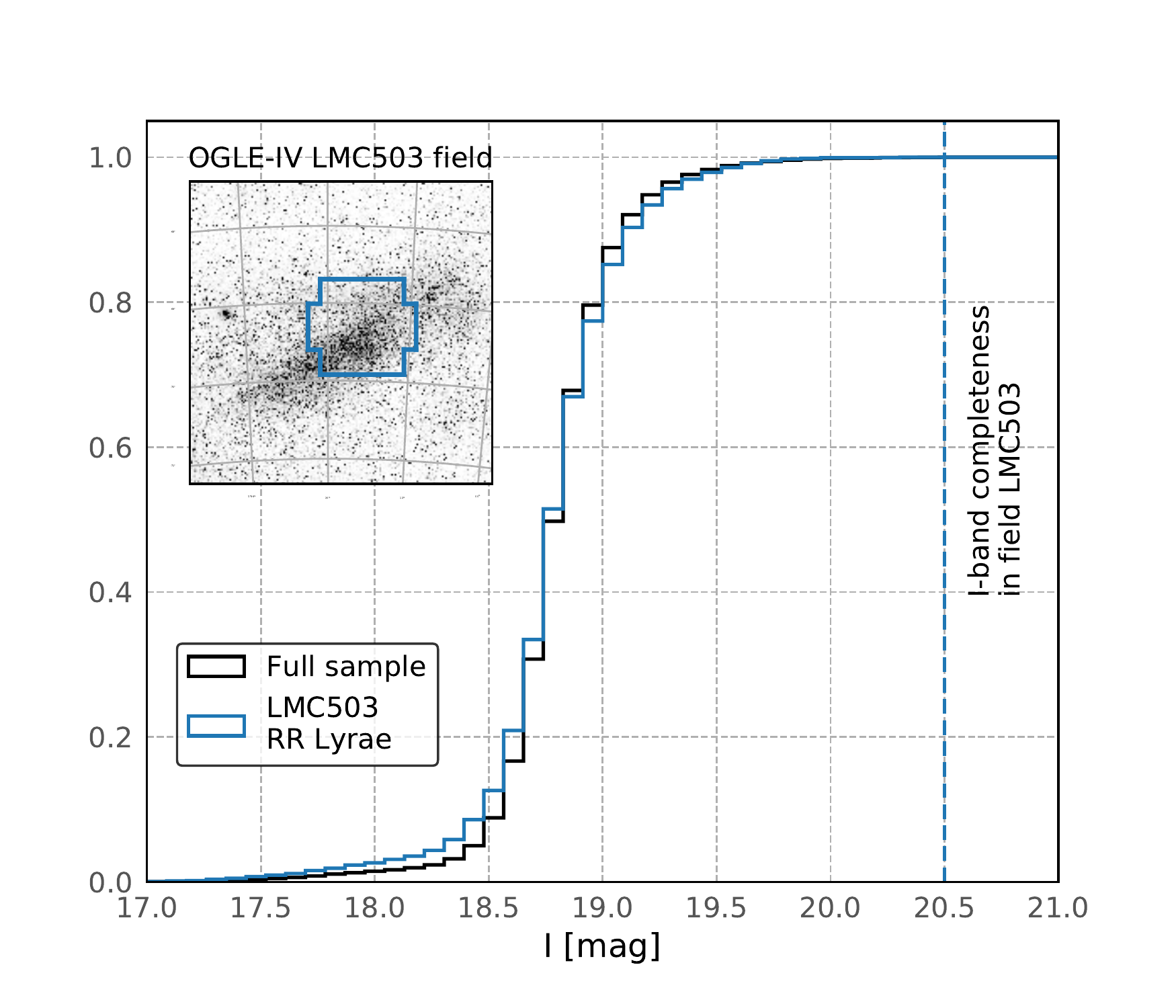}
	\caption{$I$-band luminosity function of the OGLE-IV RR Lyrae. The cumulative histograms show the fraction of RR Lyrae
          dimmer than a certain $I$-band brightness. The luminosity function of the full RR Lyrae sample from Figure \ref{fig:rrlmap} is
          shown in black, and that of RR Lyrae in the `LMC503' field of the OGLE-IV survey, a region with dense stellar crowding \citep{Udalski2015}, is shown in blue. The inset shows the location of the LMC503 region on an $r$-band continuum map (as reference) of the LMC Bar from the Magellanic Cloud Emission-Line Survey \citep[MCELS, ][]{Winkler2005, Pellegrini2012}. The $I$-band completeness limit of the LMC503 region is shown with the dashed line.}
	\label{fig:RRLcompl}
\end{figure}

In this paper, we will use a SAD map of the Large Magellanic Cloud (LMC) to  derive the DTD of RR Lyrae stars--- pulsating horizontal branch stars with periods between 0.2 and 1 day \citep{Smith2004}. We chose to test the DTD method on RR Lyrae for several reasons. Firstly, the sample size of RR Lyrae in the LMC is quite large (see Section \ref{sec:ogleivsamp}), allowing us to measure a DTD with high significance. Secondly, there is strong evidence that RR Lyrae are mostly ancient stars, older than 10 Gyrs, given their pulsational properties \citep{Smith2004, Marconi2015} and abundance in old globular clusters \citep{Clement2001, OGLEIVBULGE, OGLEIVRRL}. Measuring an RR Lyrae DTD therefore provides a rigorous test of the DTD method for the recovery of progenitor age-distributions, as well as star-formation history measurements of old resolved stellar populations. Lastly, a DTD analysis provides an opportunity to test stellar evolution models of RR Lyrae in a completely new way. While there is consensus on the interpretation of RR Lyrae as ancient stars, the lower-limit on their ages has been somewhat unconstrained.  For example, the absence of RR Lyrae in the SMC cluster Lindsey 1 ($t \approx 9$ Gyr), compared to their presence in NGC 121 ($t \approx$10--11 Gyr), is generally cited as evidence of a lower limit of 10 Gyr for progenitor age of RR Lyrae \citep{Ols1996, Glatt2008}. However, growing evidence of thin disk RR Lyrae in the Milky Way raises questions about whether this limit might be lower, and if an intermediate-age progenitor channel can exist \citep{Layden1995, Zinn2019, Prudil2020}. Additionally uncertainties may exist in the evolutionary models of RR Lyrae stars themselves: since RR Lyrae are horizontal-branch stars, their positions on the color-magnitude diagram depend on an unknown combination of factors (commonly known as the second-parameter phenomenon) like metallicity, age, mass-loss on the red-giant branch, stellar rotation, core structure, and chemical abundance \citep[see][for reviews]{Fusi1997, Catelan09,
Dotter2013}. 

The paper is organized as follows: Section \ref{sec:ingred} briefly describes the two ingredients for calculating the RR Lyrae DTD -- the OGLE-IV survey of RR Lyrae, and the \cite{Harris2009} SAD map of LMC, Section \ref{sec:rrldtd} describes the measurement of the RR Lyrae DTD from the OGLE-IV survey, Section \ref{sec:assess} checks whether the measured DTD is consistent with the observed spatial distribution of RR Lyrae in the OGLE-IV survey, incompleteness in the SAD map and RR Lyrae in star clusters, Section \ref{sec:implications} discusses the two possible interpretations of the DTD result -- that RR Lyrae may have a previously-unidentified younger progenitor channel, or unknown systematics still exist in SAD measurements of resolved stellar populations older than a Gyr.

\section{Ingredients for Calculating The DTD} \label{sec:ingred}
\subsection{OGLE-IV Sample of LMC RR Lyrae} \label{sec:ogleivsamp}
We use the \cite{OGLEIVRRL} catalog of 39,082 RR Lyrae stars from the Optical Gravitational Microlensing Experiment (OGLE-IV)\footnote{\url{http://ogle.astrouw.edu.pl/}} survey of the LMC \citep{Udalski2015}. These RR Lyrae were selected by the OGLE-IV pipeline from the full database of OGLE $I$-band light curves with periods between 0.2 and 1 day, and then further classified as fundamental (RRab), first-overtone (RRc), and mixed mode pulsators based on their periods, Fourier amplitudes, and light curve shapes. We excluded catalog entries that were flagged as Galactic RR Lyrae or eclipsing variables, objects with uncertain classification, and sources that fall outside the SAD maps. This produced a final sample of 29,810 RR Lyrae (Figure \ref{fig:rrlmap}). The photometric completeness of the sample is quite high, as evidenced by the I-band luminosity function of RR Lyrae in the most crowded OGLE-IV field (LMC503, see Figure \ref{fig:RRLcompl}). The field has a completeness limit of $I \approx 20.5$ \citep{Udalski2015}, whereas the RR Lyrae sample in the field has a median magnitude $\bar{I}=18.82$ and standard deviation $\sigma_I = 0.4$. The median magnitude of the RR Lyrae is nearly 4.2$\sigma_I$ above the completeness limit (almost 99.9$\%$ of the RR Lyrae have I-band magnitude above the completeness limit). Thus even in the most crowded region, the RR Lyrae sample can be considered photometrically complete. Also as seen in Figure \ref{fig:RRLcompl}, the I-band luminosity function of the full OGLE-IV sample inside the \cite{Harris2009} region (see Section \ref{sec:sad}) is also well above the completeness limit of the most crowded region. Such high completeness is important  for measuring unbiased rates and DTDs \citep{Maoz2010,Badenes2010}.

\subsection{Stellar Age Distribution Map of LMC} \label{sec:sad}
We use the SAD map of the LMC constructed by \cite{Harris2009} (hereafter, HZ09). This map provides the best-fit stellar mass formed as a function of lookback time in spatial cells resolving the central $8.5^{\circ} \times 7.5^{\circ}$ of the galaxy \citep{zaritsky2004a}. HZ09 also provides the associated 1$\sigma$ upper and lower limits to the SAD in each cell, which we will incorporate into the DTD uncertainties in Section \ref{sec:uncer}. The SADs were calculated using data from the Magellanic Cloud Photometric Survey (MCPS) of nearly 4 million stars collected with the 1m Swope telescope, down to a completeness of V=20--21 mag \citep{Zaritsky1997, zaritsky2004a}. The MCPS region was divided into 1376 cells, each measuring $24^{\prime} \times 24^{\prime}$, or $12^{\prime} \times 12^{\prime}$ if the field contained more than 25,000 stars. The contours of these cells are shown in Figure \ref{fig:rrlmap}. SADs were derived using the StarFISH algorithm by fitting the color magnitude diagram in each cell with a linear combination of isochrones \citep{Harris2001}. After accounting for extinction and photometric errors, each cell's SAD was fit using 16 logarithmically-spaced bins spanning the ages between 4 Myr and 20 Gyr, and four metallicity bins ($Z=0.008$, 0.004, 0.0025, and 0.001). For ages younger than 100 Myr, a single metallicity of Z=0.008 was used because the different metallicity isochrones are almost indistinguishable.

Note that while the lower limit on the age of RR Lyrae stars is generally quoted as 10 Gyr, the  SAD  map  has  a  single  indivisible  age-bin  of  8--12 Gyr, and so we refer to this lower limit as 8 Gyr in the rest of the paper.

\section{The RR Lyrae Delay-Time Distribution} \label{sec:rrldtd}

\subsection{Method} \label{subsec:method}

The RR Lyrae catalog and SAD maps of the LMC are used to estimate the RR Lyrae DTD using the non-parametric method described in \cite{Badenes15} (hereafter, B15), although we improve on some aspects of
it. The RR Lyrae DTD is the number of RR Lyrae formed per unit stellar mass as a function of the time-delay between star-formation and the RR Lyrae phase.
The convolution of the DTD with the SAD in each spatial cell predicts the number of RR Lyrae that will be produced by the stellar population in that cell. 
For each SAD cell $i$, the
number of RR Lyrae predicted ($\lambda_i$) is:
\begin{equation} \label{eq:lambda_i}
	\lambda_i = \sum_{j=1}^{N} M_{ij} \left(\Psi T_{vis}\right)_j
\end{equation}
where $M_{ij}$ is the stellar mass formed in age-bin $j$ and cell $i$, $\Psi$ is the RR Lyrae formation rate (RR Lyrae per year per unit M$_{\odot}$), and T$_{vis}$ is the duration of the RR Lyrae phase (both $\Psi$ and T$_{vis}$ are function of age-bin). The widths of the age-bins, $j$, are
selected using the methodology described in Appendix \ref{sec:binning} and provide the best compromise between detection
significance and temporal resolution. 

In this paper, we retain the notation of B15 and refer to the quantity $\left(\Psi T_{vis}\right)$ as the DTD (with units of RR Lyrae per unit M$_{\odot}$). The DTD ($\Psi T_{vis}$) is determined by minimizing the difference between the predicted and observed number of RR Lyrae across spatial cells. This is carried out in a similar manner to B15 using the Markov Chain Monte Carlo (MCMC) solver
\texttt{emcee} \citep{emcee}. We denote $\mathbf{N} = [N_i]$ as the
vector representing the number of RR Lyrae in each spatial cell, and $\mathbf{\Psi T_{vis}} = [\left(\Psi T_{vis}\right)_j]$ as the vector of
predicted number of RR Lyrae per stellar mass for each age-bin. The posterior is calculated as
\begin{equation}
	p\left(\mathbf{\Psi T_{vis}} | \mathbf{N}\right) \propto \mathcal{L}\left(\mathbf{N} | \mathbf{\Psi T_{vis}}\right) \pi\left(\mathbf{\Psi T_{vis}}\right)
\end{equation}
where $\mathcal{L}$ is the likelihood and $\pi$ is the prior.
We assume $\mathcal{L}$ is either a product of Poisson or Gaussian probabilities depending on N$_i$,
 \begin{equation}
\mathcal{L}\left(\mathbf{N} | \mathbf{\Psi T_{vis}}\right) =
 	 \begin{cases}
    		\displaystyle \prod_{i=1}^{K} \frac{\mathrm{e}^{-\lambda_i} \lambda_{i}^{N_i}}{N_i !}      & \quad N_i \leq 25\\
    		\displaystyle \prod_{i=1}^{K} \frac{1}{2\pi \lambda_i} \left(\frac{(\lambda_i - N_i)^2}{2\lambda_i}\right)  & \quad N_i > 25
 	 \end{cases}
	 \label{eq:lik}
\end{equation}
where $K$ is the number of SAD cells. Although Poisson distributions converge to Gaussian for large $N_i$, we specify the likelihoods separately to avoid computational
issues. The prior is defined such that the $(\Psi T_{vis})_j$ values have uniform probabilities in log-space:
\begin{equation}
	\pi \left(\mathbf{\Psi T_{vis}}\right) = 
		\begin{cases}
			\mathbf{\left(\Psi T_{vis}\right)^{-1}} & \quad (\Psi T_{vis})_j > 0\\
			0 						      & \quad (\Psi T_{vis})_j \leq 0
		\end{cases}
\end{equation}

\subsection{Estimating Uncertainties} \label{sec:uncer}

We propagate the errors on the $M_{ij}$ values in the SAD map into uncertainties on $\mathbf{\Psi T_{vis}}$. 
B15 calculated DTDs for the best-fit SAD, the upper-limit on the SAD, and the lower-limit on the SAD. The differences between the best-fit and upper/lower limits were treated as the DTD's 1$\sigma$  uncertainties. 

In this paper, we use an improved method of propagating SAD uncertainties into the DTD. We randomly generate 100 mock SADs, assuming $M_{ij}$ has the
normally-distributed uncertainties given by HZ09, and calculate a DTD for each mock SAD\null. We combine the MCMC posterior chains from these 100 DTDs into
a single chain, and estimate the $95\%$ credible interval on this chain using a highest posterior density criterion. We define the mode of the distribution minus the upper and lower-limits of this interval as our $2\sigma_{+}$ and $2\sigma_{-}$ confidence intervals, respectively. We define a ``signal" detection in each bin $j$ of the DTD as a value of $\Psi T_{vis}$ in that bin that is $\geq$ $2\sigma_{-}$ above 0. Non-detections are presented as $2\sigma$ limits on the DTD signal in a particular age bin. 

\subsection{Sample Contribution per Age-bin}
We estimate the contribution of each age-bin to the total sample by multiplying the value of $\Psi T_{vis}$ with the total stellar mass formed in each age-bin. The contribution percentage will then have uncertainties due to both the DTD and the total stellar masses formed. We estimate these uncertainties with a Monte Carlo method. We draw DTD values from the recovered posterior probability distributions (called $\left(\Psi T_{vis}\right)_j$), multiply by the
total stellar mass per age-bin in each randomized SAD map (called M$_{j}$), and get the number of RR Lyrae contributed by
age-bin, $\lambda_j$. The contribution percentages (= $\lambda_j/\Sigma \lambda_j$) and their 1$\sigma$ uncertainties are
listed in Table \ref{tab:rr}.

\begin{figure}
	\includegraphics[width=\columnwidth]{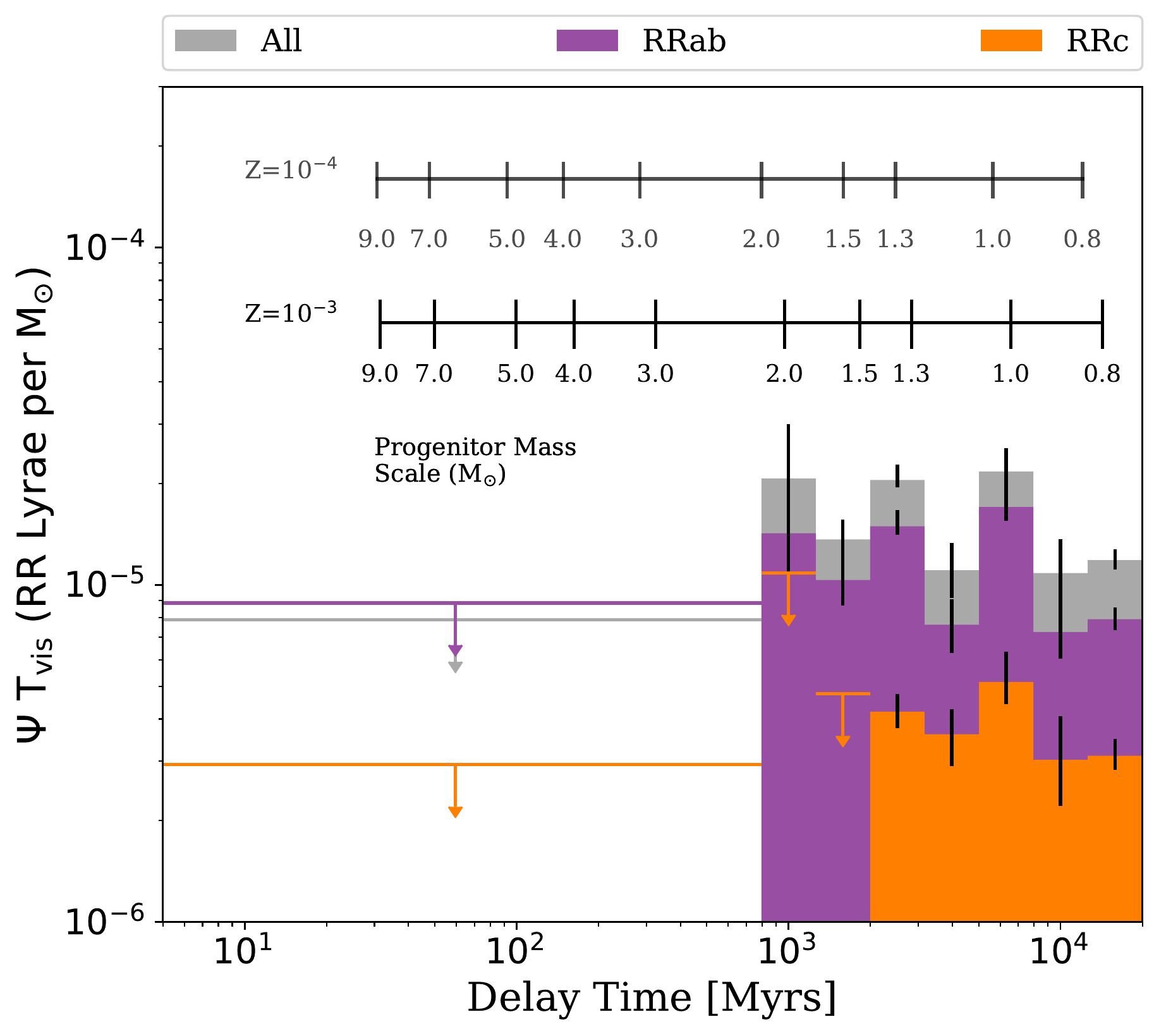}
	\caption{The delay-time distribution, in units of number of RR Lyrae per M$_{\odot}$, as a function of delay-time in Myr. The grey histogram represents the DTD for the full OGLE-IV sample, while the colored histograms are DTDs for RR Lyrae subtypes RRab and RRc
	Filled histograms represent bins with a statistically significant signal from the MCMC analysis. The error bars are 1$\sigma$
    uncertainties that include uncertainties from the SAD maps. The arrows represent 2$\sigma$ upper limits in bins that do not have a statistically significant signal. At top, the progenitor mass-scales for two different metallicities, calculated using PARSEC, are shown for reference.}
	\label{fig:dtd}
\end{figure}
      
\subsection{Results} \label{subsec:results}
      
The DTD for the full OGLE-IV RR Lyrae sample and the two main RR Lyrae subtypes is shown in Figure \ref{fig:dtd}, and the values are tabulated in Table \ref{tab:rr}. We detect signal in the DTD at a high significance ($> 5\sigma$) for all age-bins older than 1.3 Gyrs. The detections with the highest significance are found in the 2--3, 5--8 and $>12$ Gyr bins. Most of this signal is contributed by the RRab stars, which are the most common subtype in the OGLE-IV sample. The RRc subclass contributes to the DTD mostly above 2 Gyr. Although RRc's have been susceptible to confusion with other variable sources in time domain-surveys \citep{Kinman2010, Mateu2012, Drake2014}, Figure \ref{fig:dtd} shows that the total DTD is dominated by RRab objects in all age-bins, making it unlikely that the DTD is biased by sample contamination. The detected signal of the full DTD is relatively flat above 1.2 Gyr, with about $1-3$ RR Lyrae produced per 10$^{5}$ M$_{\odot}$ of stellar mass formed. While there is signal in the DTD below 0.8 Gyrs, it
falls below our 2$\sigma_{-}$ threshold, and we only show the 2$\sigma_{+}$ upper limit.

Our detection of a DTD signal below 8 Gyrs is a surprising result. About $46\%$ of the LMC's RR Lyrae stars are produced from populations older than 10 Gyr, the age range generally inferred for RR Lyrae in star clusters. But about $51\%$ of the DTD signal comes from progenitors with ages between 1.3 and 8 Gyr, and this result has high ($>5\sigma$) significance.  If we re-calculate the DTD without assuming normal errors on $M_{ij}$ (i.e., using the same method as B15, see Appendix \ref{app:b15}), we still detect a strong signal below 8 Gyr, but with a total contribution of 41$\%$. A comparison of these timescales with those of the PARSEC\footnote{\url{https://people.sissa.it/~sbressan/parsec.html}. We use PARSEC because it is one of the latest stellar evolution codes with both publicly available main-sequence and horizontal branch tracks. We get similar results with the MIST evolutionary tracks of \cite{Choi2016}.} models for the onset of helium burning leads to the conclusion that RR Lyrae can arise from main-sequence progenitors as massive as $\sim 2$ M$_{\odot}$ at LMC metallicity. Incidentally, this upper limit is similar to the mass at which stars transition between igniting He under degenerate conditions (the ``He-flash'') to burning helium under stable, non-degenerate conditions \citep{Bildsten2012, Mosser2014}. We also tested the dependence of the DTD as a function of RR Lyrae star period and brightness by sub-dividing the OGLE-IV data by pulsation time (3 bins with $<0.45$ d, $0.45-0.58$ d, and $>0.58$ d) and $I$-band magnitude (3 bins with I$<19.2$, $19.2-19.4$ and $>19.4$).  By sub-dividing the sample in this way, we ensured that each period and magnitude bin contained enough RR Lyrae for a robust measurement of their DTDs. We found that all the sub-samples have significant ($>2\sigma$) DTD signals in the range 1 to 8 Gyr, with no discernible trend in the shape of the DTD with magnitude or period.

\begin{figure*}
    \subfigure[]{\includegraphics[width=0.5\textwidth]{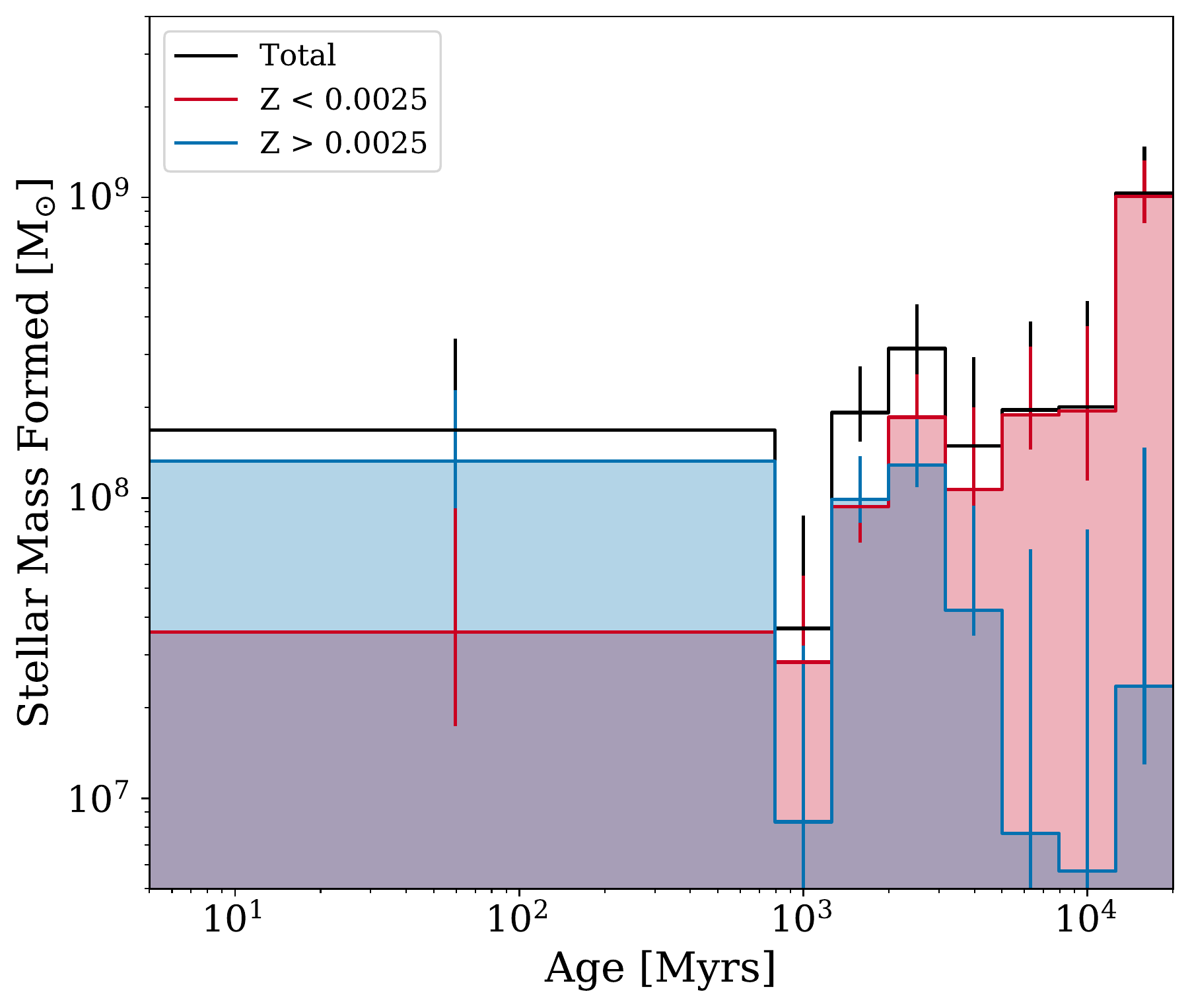}\label{fig:sadmetallicity}}
    \subfigure[]{\includegraphics[width=0.51\textwidth]{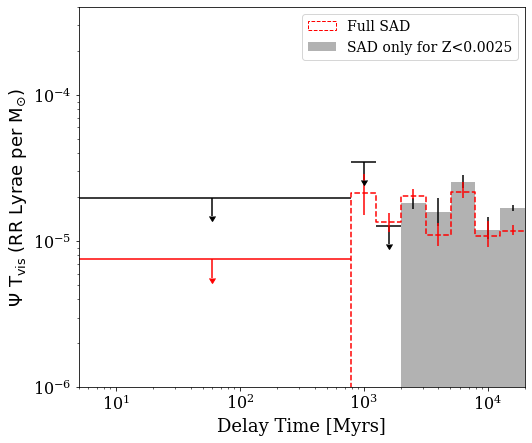}\label{fig:dtdmetallicity}}
    \caption{(a) The global (summed over all SAD cells) stellar mass formed versus age ($M_{ij}$ in Eq \eqref{eq:lambda_i}) with $Z<0.0025$ (red) and $Z>0.0025$ (blue) in the LMC as given by the HZ09 SAD. The black histogram shows the sum of the red and blue histograms. Overlapping regions appear in a darker shade. (b) The DTD measured from the OGLE-IV survey using only the SAD in the $Z<0.0025$ bin (grey; see Section \ref{subsec:results} for details). For comparison, the DTD from Figure \ref{fig:dtd} is shown in red.}
\end{figure*}

Although, in principle, it is possible to derive a metallicity-dependent DTD with the HZ09 SAD maps, we defer that study to a future work.  However, it is well-known that LMC RR Lyrae are generally metal-poor, with [Fe/H]$<-0.5$ and a peak in the metallicity distribution function at [Fe/H]$\sim -1.5$ \citep{Skowron2016}.  We can therefore check to see if there is DTD signal below 8 Gyrs under the assumption that OGLE-IV RR Lyrae are only produced by the HZ09 SAD in the two metal-poor bins, i.e., for $Z<0.0025$ or [Fe/H]$\lesssim-1.02$ \citep{Bertelli1994}. Star-formation in this metallicity range dominated the LMC SAD until $\sim 2$ Gyrs, as seen in Figure \ref{fig:sadmetallicity}, and the resulting DTD is shown in Figure \ref{fig:dtdmetallicity}. Even if we restrict our analysis to the metal-poor SAD, the signal below 8 Gyrs persists. This implies that, even if we assume that all LMC RR Lyrae are produced by metal-poor stars, progenitors younger than 8 Gyrs are still needed to explain their distribution. 

Before delving into the implications for RR Lyrae studies, we carry out a more detailed analysis of the recovered
DTD to assess the robustness of our results.

\begin{deluxetable}{lccc}
\tablecaption{The RR Lyrae DTD calculations, with lifetimes, significance of detection and contribution to the OGLE-IV RR Lyrae sample considered in this study.\label{tab:rr}}
\tablehead{\colhead{\textbf{Delay-Times}} & \colhead{DTD ($\Psi T_{\rm{vis}}$)} & \colhead{Significance} & \colhead{Contribution}\\
\colhead{(Gyr)} & \colhead{(N/10$^5$ M$_{\odot}$)} & \colhead{($\sigma$)} & \colhead{( $\%$)}}
\startdata
$<$ 0.8 & $< 0.75$ & $-$ & $-$\\
0.8--1.3 & $2.12^{+0.61}_{-0.78}$ & $3.8$ & $2.9\pm0.8$\\
1.3--2.0 & $1.36^{+0.2}_{-0.21}$ & $5.6$ & $8.4\pm1.2$\\
2.0--3.2 & $2.05^{+0.11}_{-0.22}$ & $20.5$ & $20.7\pm1.2$\\
3.2--5.0  & $1.10^{+0.18}_{-0.23}$ & $5.6$ & $6.0\pm1.0$\\
5.0--7.9 & $2.17^{+0.19}_{-0.36}$ & $11.6$ & $15.6\pm1.3$\\
7.9--12.6 & $1.08^{+0.17}_{-0.29}$ & $6.8$ & $7.5\pm1.1$\\
12.6--20.0 & $1.18^{+0.07}_{-0.1}$ & $18.5$ & $39.0\pm1.8$\\
\enddata
\end{deluxetable}

\section{On the robustness of the recovered RR Lyrae DTD} \label{sec:assess}
\subsection{Mock DTD Test} \label{sec:mockdtd}
The HZ09 SAD map of the LMC allows us to test whether the traditional DTD for RR Lyrae (i.e., one in which the progenitors are always older than 8 Gyr) is consistent with the spatial distribution of RR Lyrae observed by OGLE\null. This is equivalent to inverting the DTD recovery process. To do this, we generate mock RR Lyrae maps by convolving the SAD map with a DTD that is non-zero only in the 2 oldest
age-bins, 8--12 and 12--20 Gyr. The total stellar mass that formed in these age-bins was $1.23 \times 10^{9}$ M$_{\odot}$. 

Assuming all 29,810 RR Lyrae were produced by progenitors in these age-bins results in a DTD of the form,
\begin{equation} \label{eq:fakedtd}
	\Psi T_{vis} = 
		\begin{cases}
			2.42 \times 10^{-5} \mathrm{\ RRL\ M_{\odot}^{-1}} & \quad t \geq 8\ \mathrm{Gyrs} \\
			0 						      & \quad t < 8\  \mathrm{Gyrs}
		\end{cases}
\end{equation}
We generate 100 mock RR Lyrae maps using this DTD, where the number of RR Lyrae per cell, $N_i$ is drawn from the Poisson distribution in Equation \eqref{eq:lik} with $\lambda_i$ given by Equation \eqref{eq:lambda_i}.

\begin{figure}

	\includegraphics[width=\columnwidth]{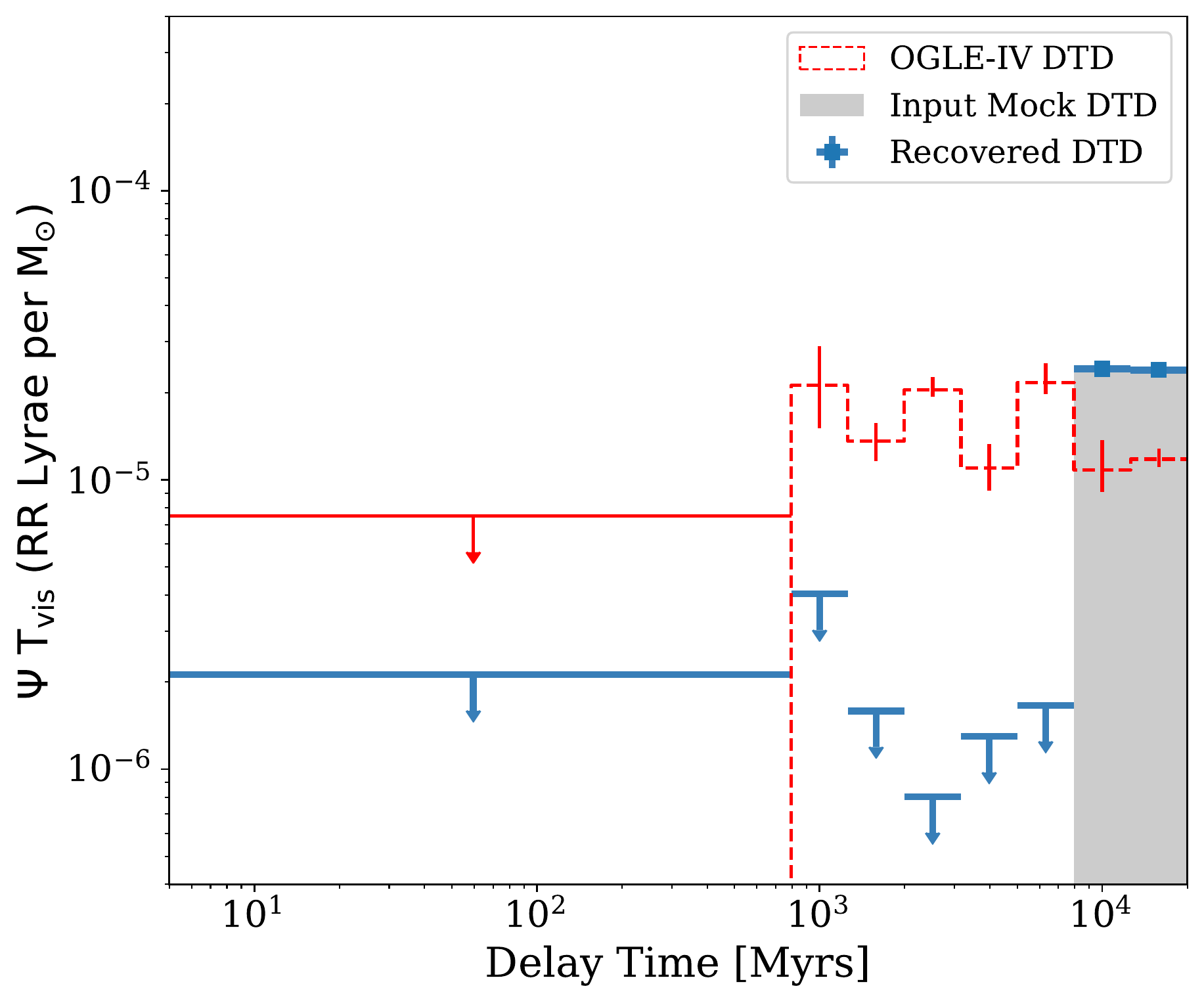}
	\caption{Result of our mock DTD test in Section \ref{sec:mockdtd}. The red histogram shows the DTD measured with the original OGLE-IV survey in Figure \ref{fig:dtd}. The black histogram shows the mock input DTD defined in Eq \ref{eq:fakedtd}. The blue histogram with arrows shows the DTD recovered from the mock input DTD and the SAD maps of HZ09.}
	\label{fig:fakedtd}
\end{figure}

\begin{figure*}
	\includegraphics[width=\textwidth]{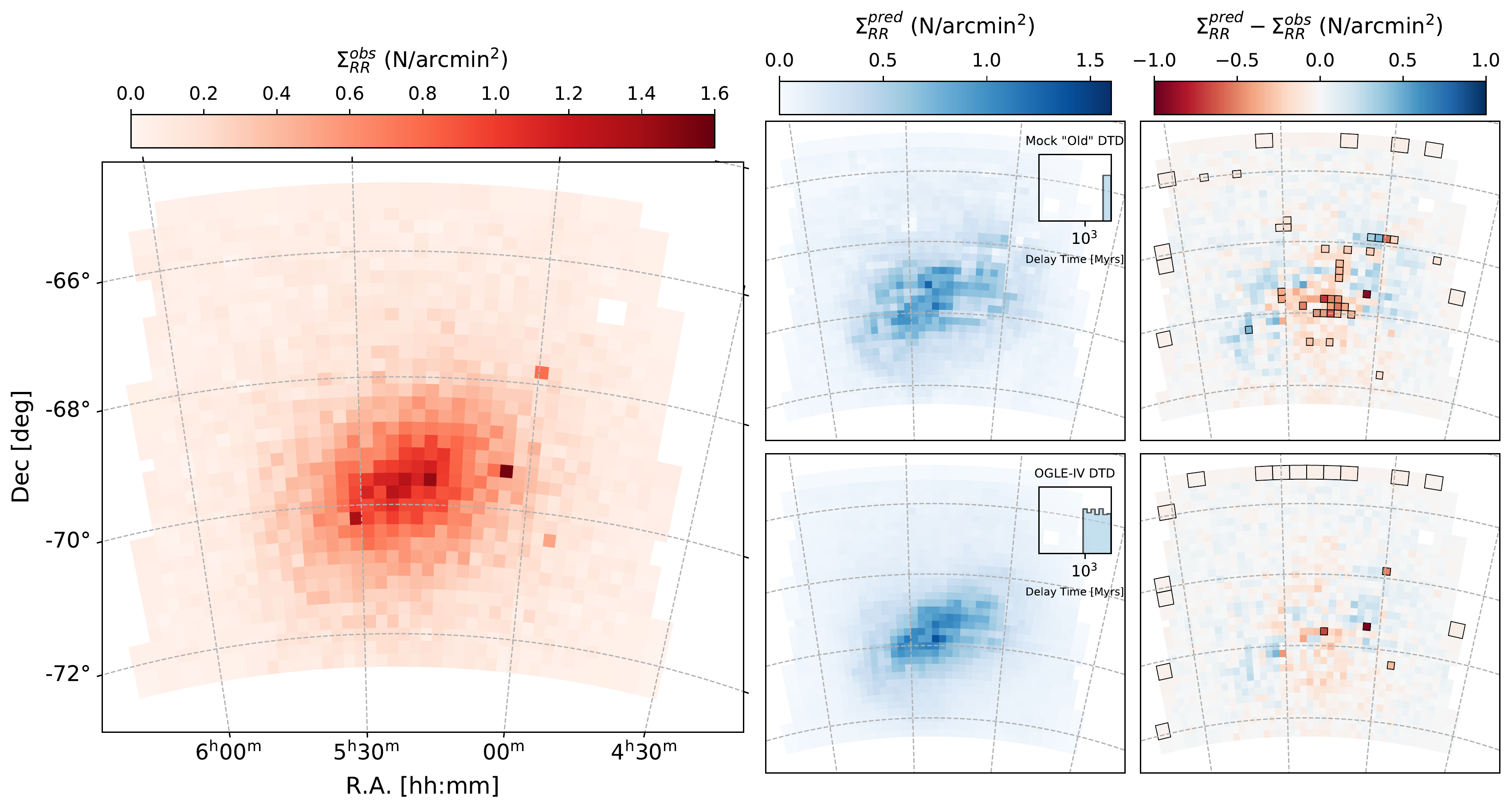}
	\caption{Comparison between the spatial distribution of RR Lyrae in OGLE-IV (left, large panel) and the distributions predicted by convolving the HZ09 SAD with the ‘mock’ old DTD (upper middle panel) and our recovered RR Lyrae DTD (lower middle panel).  The right panels show ``residuals", or the difference between the observed distribution and DTD-predicted distribution of RR Lyrae for the ‘mock’ old DTD (top right) and measured DTD (bottom right). The black squares show the cells where the difference between predicted and observed number of RR Lyrae differ by 5$\sigma$ (see Section \ref{sec:mockdtd} for details)}. 
        
	\label{fig:3panel}
\end{figure*}

Figure \ref{fig:fakedtd} shows the DTD recovered from this analysis. Our MCMC solver correctly measures the input mock DTD with strong detections only in the two oldest age-bins. The younger bins have 2$\sigma$ upper limits that are almost an order of magnitude lower than the DTD recovered from the OGLE-IV sample.  Moreover, the difference between populations of RR Lyrae with the OGLE-IV and mock DTDs can be seen visually in Figure \ref{fig:3panel}. The measured OGLE-IV DTD predicts a distribution of RR Lyrae stars elongated along the LMC Bar and declining smoothly with radius (center bottom panel of Figure \ref{fig:3panel}); this is very similar to what is seen in the OGLE-IV map (left panel).  In contrast, our mock distribution from a uniformly old RR Lyrae progenitor population (Equation \eqref{eq:fakedtd}) has more structure and leaves larger residuals in the difference map (top center and right panels of Figure \ref{fig:3panel}). To identify significant regions of discrepancy in the OGLE-IV DTD and mock old DTD maps, we generate 10$^4$ maps of RR Lyrae, where in each map the number of RR Lyrae in each cell is drawn from a Poisson distribution with mean number of RR Lyrae in that cell given by the DTD according to Eq \ref{eq:lambda_i}. We then estimate the mean and standard deviation of the number of RR Lyrae per cell, and identify cells with black squares in Figure \ref{fig:3panel} where the observed RR Lyrae is greater than 5 times the standard deviation from the mean. We find the mock old DTD map has a larger number of cells where the RR Lyrae number count is discrepant, and these cells are mostly located in the Bar region, with a few located outside (note that this is only for the purpose of visualizing the discrepancy; the actual DTD is constrained by the joint likelihood measured from \emph{all} the cells as in Eq \ref{eq:lik}.) 

We conclude that if the OGLE-IV RR Lyrae had indeed been produced exclusively from the old ($t > 8$ Gyr) stars of the HZ09 SAD map, our method would have recovered the correct DTD, without spurious signals at younger ages. The DTD signal we recover at ages between 0.8 and 8 Gyr must either be real, or an artifact produced by systematics in the HZ09 SADs. Next, we investigate the impact of these SAD systematics on the DTD more thoroughly.

\subsection{Effect of incompleteness and crowding in the SAD map} \label{sec:sadcompl}
We explored if any incompleteness in the MCPS photometry (from which the SAD was measured) could be driving the intermediate-age signal. The photometric completeness limit is around V=21 mag, with the Bar region being almost 1 mag shallower as a result of the stellar crowding. Because of this, StarFISH could reliably solve for the number of stars per age-bin only for ages younger than 4 Gyr (although we note that our intermediate-age DTD signal is detected younger than 4 Gyrs). Populations older than 4 Gyr were traced by their giant branch stars, and assigned by StarFISH to a single age-bin.
 To anchor the `shape' of the SAD beyond 4 Gyrs, stellar ages were measured in a few isolated \emph{HST}-fields in the LMC Bar \citep{Holtzman1999, Olsen1999, Smecker2002}. As shown in Figure 7 of HZ09, the SADs of each \emph{HST} Bar field are characterized by star-formation at look-back times of 10 Gyrs and 5 Gyrs, with a period of quiescence in between.  This consistency allowed HZ09 to adopt a common shape for the SAD in the rest of the Bar region.

\begin{figure}
	\includegraphics[width=\columnwidth]{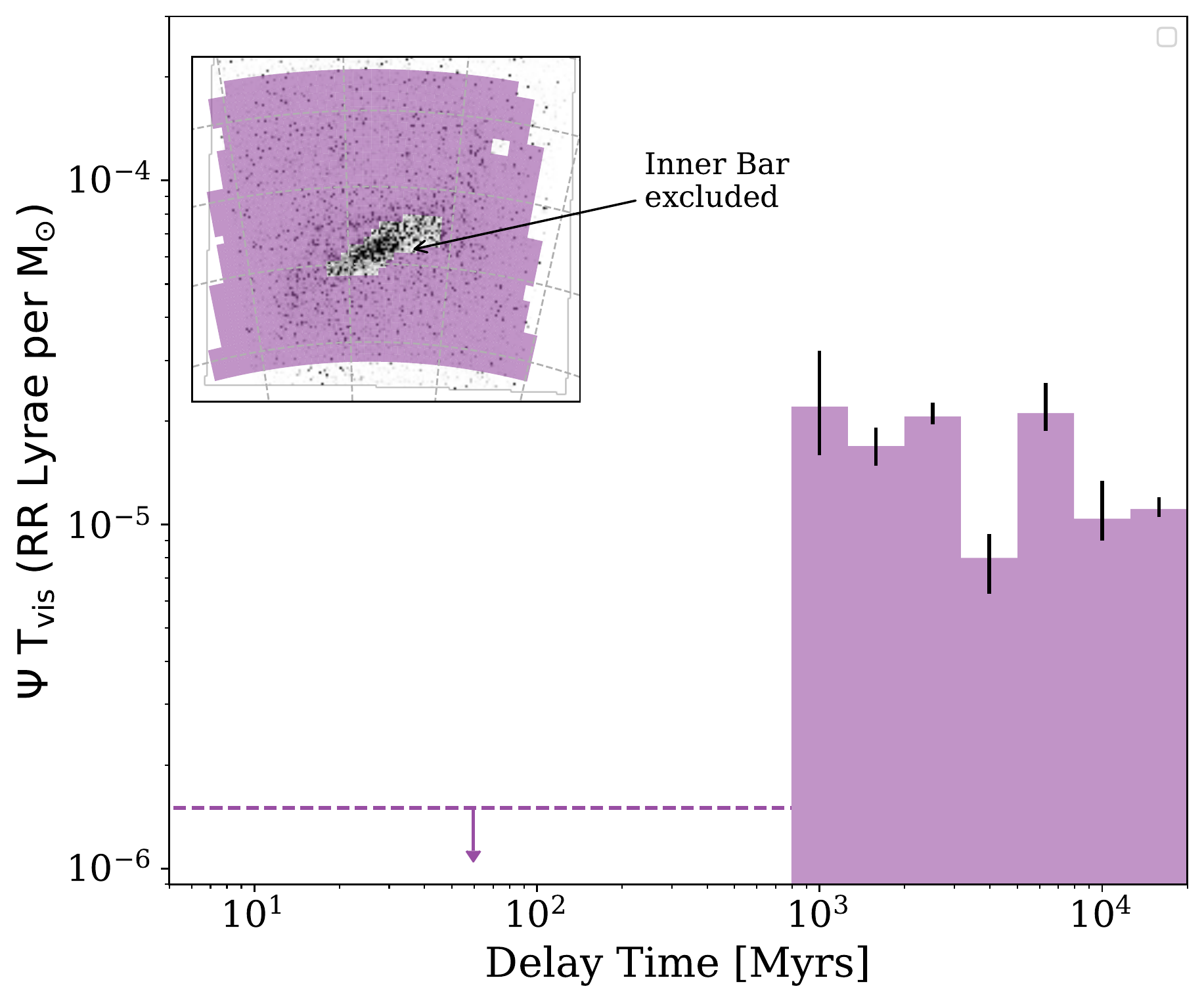}
	\caption{RR Lyrae DTD of the OGLE-IV survey, but excluding the SAD cells and RR Lyrae of the Inner Bar. The excluded cells of HZ09 are shown in the inset plot, overlaid on the $r$-band continuum map from MCELS.}
	\label{fig:dtdnobar}
\end{figure}

\begin{figure}
	\includegraphics[width=\columnwidth]{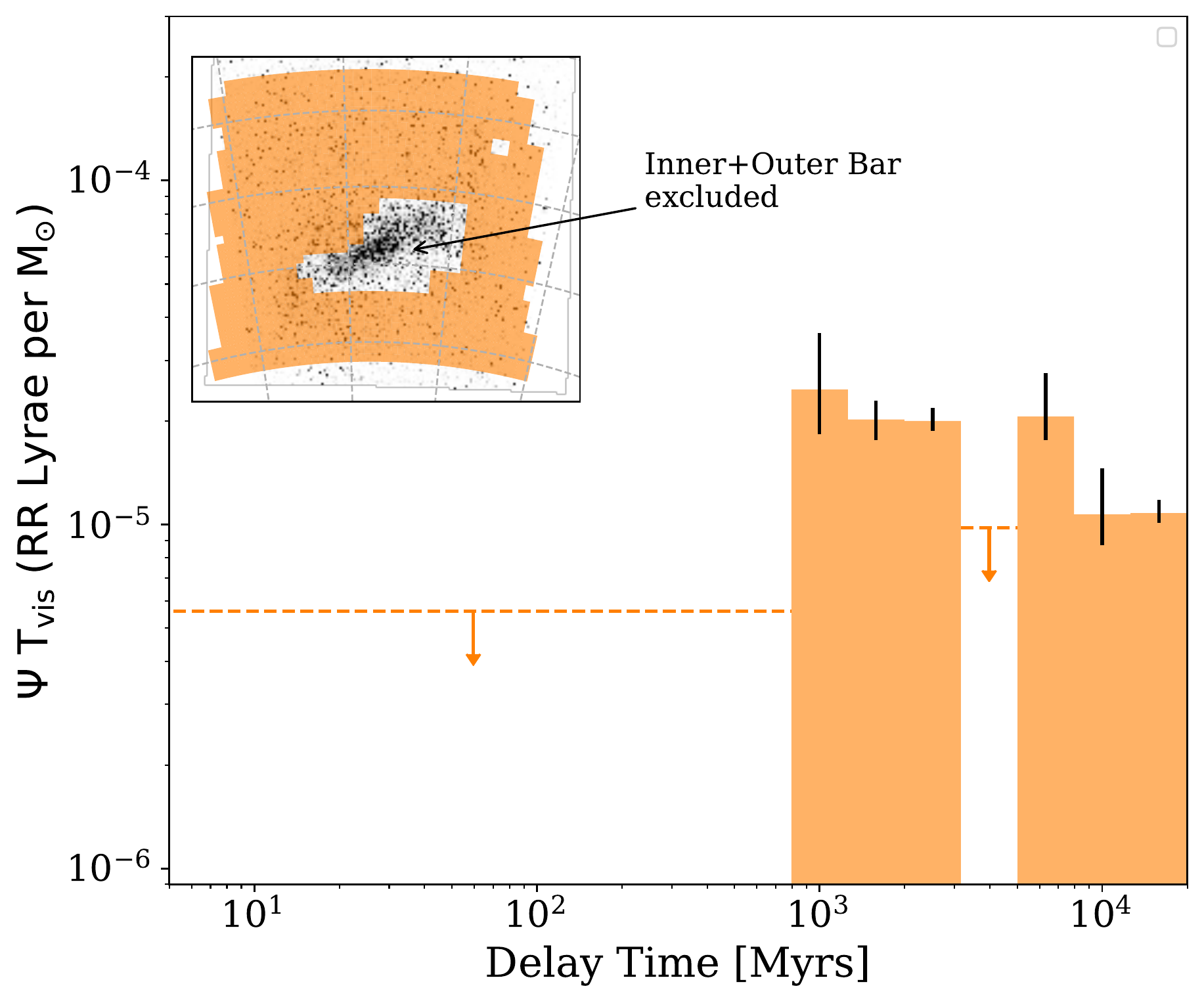}
	\caption{Same as Figure \ref{fig:dtdnobar}, but now excluding both the Inner and Outer Bar. Note that the excluded cells for both Inner and Outer Bar are based on Figure 6 of HZ09.}
	\label{fig:dtdnoallbar}
\end{figure}

We first checked if an underestimation of the total mass formed at old stellar ages due to any photometric incompleteness could be driving the intermediate-age signal. Obtaining a complete census of old stellar mass is generally difficult without including infra-red data \citep{Conroy2013}, so we studied the changes to our DTD by manually changing the old SADs. We recalculated the RR Lyrae DTD with the same SAD map, except with each cell's 8--12 Gyr and 12--20 Gyr stellar mass multiplied by factors of 2, 4 and 10. We find that regardless of how much mass is added, the DTDs still find signals for ages between 0.8 and 8 Gyrs with $>5\sigma$ confidence. Since the incompleteness in the SAD map is dominant in the Bar region, we also tried re-calculating the DTD without the LMC Bar (i.e., by removing the SAD cells and their RR Lyrae from our
calculations). We show the RR Lyrae DTDs without the ``Inner Bar'' in Figure \ref{fig:dtdnobar}, and without both the ``Inner'' and
``Outer Bar'' in Figure \ref{fig:dtdnoallbar} (the cells of these regions are defined according to Figure 6 in HZ09). The recovered DTDs outside the excluded regions are similar to the DTD obtained from the full OGLE-IV sample. For the case with both the Inner and Outer Bar removed, the number of RR Lyrae and the number of SAD cells are smaller, leading to a higher upper limit on the DTD older than $0.8$ Gyr, and a non-detection in the 3--5 Gyr bin. However, there is still significant signal at ages younger than 8 Gyr, and particularly below 4 Gyr where the ages are determined by the main-sequence turnoff in the MCPS photometry. It therefore appears unlikely that missing stellar masses at ages $>8$ Gyrs due to incompleteness, or crowding in the Bar region are the only factors driving the intermediate-age signal in the DTD.


\subsection{Comparison with RR Lyrae in LMC star clusters} \label{sec:clusters}
\begin{figure*}
    \centering
    \includegraphics[width=\textwidth]{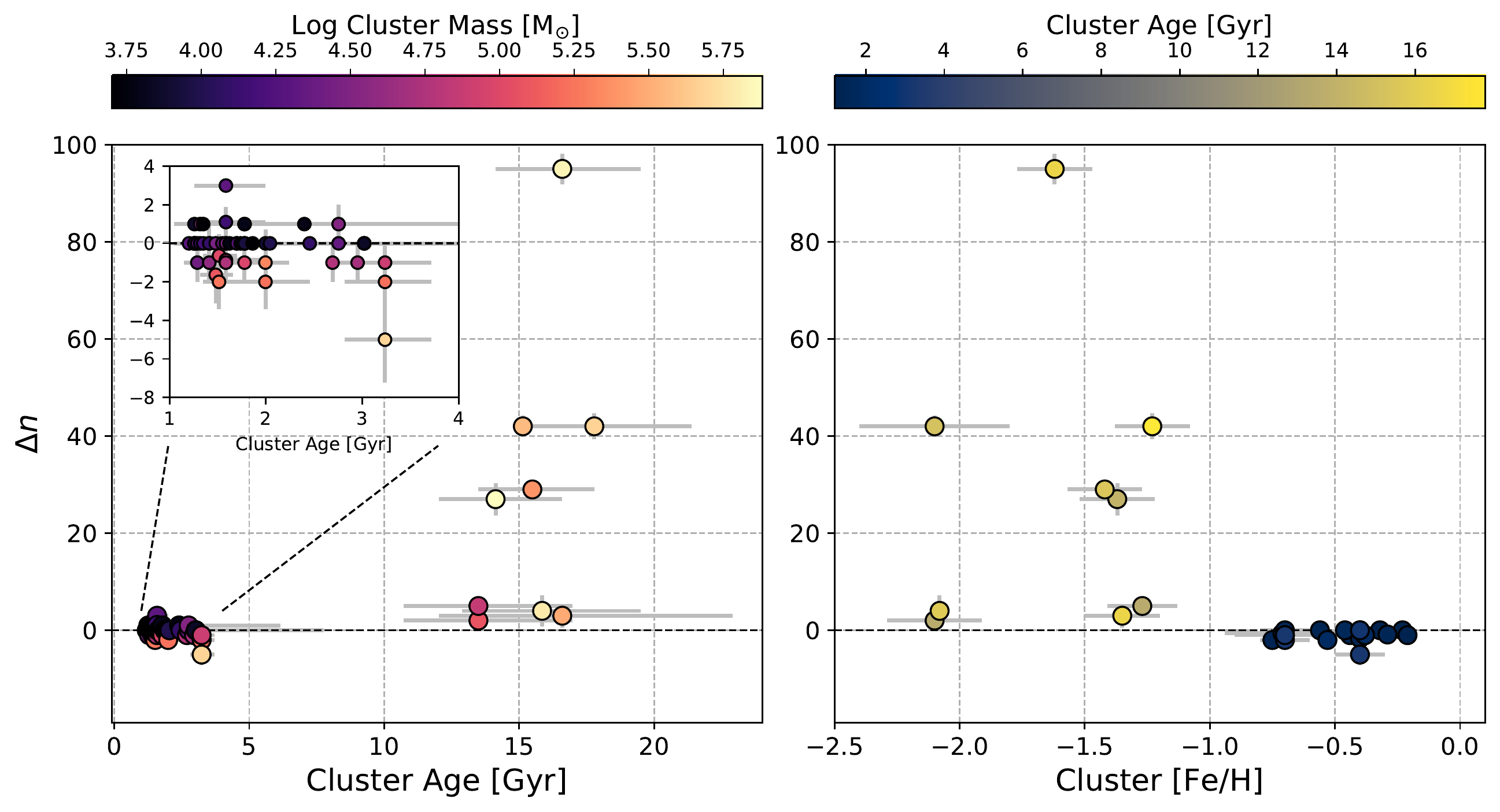}
    \caption{(\emph{Left}): Comparison of ages of the 62 clusters in \cite{Baumgardt2013} older than 1 Gyr, and their $\Delta n = (n_o - n_b) - n_p$, the difference between the background-subtracted number of RR Lyrae observed per cluster, and the number predicted by the DTD given the cluster mass and age (see Eq \eqref{eq:nb}, \eqref{eq:n_p} and Section \ref{sec:clusters} for details). Colors indicate the cluster mass. The inset zooms-in on clusters between 1--4 Gyr. (\emph{Right}): $\Delta n$ vs metallicity for a subset of 29 clusters with metallicity information available in the literature. Colorbar indicates the cluster age. The inset zooms-in on the 1--4 Gyr clusters, which also happen to cluster in the metallicity range $-1 \lesssim$ [Fe/H] $\lesssim -0.25$.}
    \label{fig:clusters}
\end{figure*}
Our measured DTD suggests that progenitors of RR Lyrae can be as young as $\sim$1 Gyr, which is much smaller than the lower limit on RR Lyrae age of 10 Gyrs inferred from star clusters \citep{Ols1996}. We therefore examine the number of OGLE-IV RR Lyrae in the LMC star clusters as a function of age, realizing that cluster membership can only be confirmed with spectroscopic and proper motion measurements that are beyond the scope of this work. 

To perform this test, we use the catalog of LMC clusters compiled by \cite{Baumgardt2013} and only include the 296 systems inside the HZ09 area. \cite{Baumgardt2013} compiled ages and masses for these clusters measured using either isochrone fitting or broadband spectral energy distribution fitting of data obtained by previous surveys \citep{Pietrzynski2000, Hunter2003, Mackey2003, deGrijs2006, Milone2009, Glatt2010, Popescu2012}. We determine the cluster radii, $r_c$ from their major ($a$) and minor axes ($b$) reported in \cite{Bica2008} as $r_c = (a + b)/4$. \cite{Bica2008} notes that these are `apparent' sizes (measured as far as the background limit in the images), but they most likely enclose the majority of the cluster mass. We also obtained chemical abundances for 29 clusters ([Fe/H], shown in Table \ref{tab:clusters} with references), and as we show later, this is sufficient for understanding the correlation of cluster RR Lyrae statistics with metallicity. Although there have been more recent studies of the LMC star cluster population \citep[e.g.][]{Nayak2016, Bitsakis2017, Piatti2017, Piatti2018}, the effects of field star contamination and asterisms on cluster identification in these studies are unclear. Since \cite{Baumgardt2013} compiles previously identified clusters above 5000 M$_{\odot}$, we rely on this catalog for our analysis. Additionally, this catalog (to the best of the authors’ knowledge) is the only catalog of LMC clusters in the literature with both age \emph{and} mass estimates of the clusters---both being critical to our analysis.  

Out of the 296 clusters, we investigate the RR Lyrae statistics of 62 clusters that have ages above 1 Gyr in order to compare with the DTD. We define $n_o$ as the number of OGLE-IV RR Lyrae observed within a circle of radius $r_c$ centered on the cluster, and $n_b$ as the expected number of RR Lyrae within the cluster area but unassociated with the cluster (we refer to these as `background’ RR Lyrae). Assuming $N$ is the number of RR Lyrae within radii $2 r_c$ and $4 r_c$ of the cluster (we allow a buffer region of $r_c$ to account for uncertainty in the actual extent of the cluster), we define $n_b$ as,
\begin{equation}
    n_b = \frac{N \pi r_c^2}{\pi(4r_c)^2 - \pi (2 r_c)^2} = \frac{N}{12}
    \label{eq:nb}
\end{equation}
Values for $n_o$ and $n_b$ are listed in Table \ref{tab:clusters} and shown in Figure \ref{fig:clusters}. 

LMC globular clusters older than 10 Gyr clearly host populations of multiple RR Lyrae that exceed the background. The richest population is in NGC 1835 with over a 100 RR Lyrae, while NGC 2005, NGC 1928 and NGC 1939 have less than 10 RR Lyrae. In contrast the intermediate-age clusters (1--10 Gyr old) are generally lacking in RR Lyrae. Out of the 53 intermediate-age clusters, 42 have 0 RR Lyrae and only three clusters (NGC 1978, SL180 and HS190) have more than 1 RR Lyrae star after subtracting the background numbers (i.e. $n_o - n_b$). This is also the case in the Milky Way, where studies of variable stars in well known intermediate-age systems such as M67 \citep{m67vars}, NGC 188 \citep{ngc188vars}, NGC 6791 \citep{ngc6791vars} and NGC 6253 \citep{ngc6253vars} have found no RR Lyrae that are likely to be cluster members.  In contrast, the Milky Way's globular cluster system contains almost 2000 of such objects \citep{Clement2001, OGLEIVBULGE}.

While this deficit of RR Lyrae in intermediate-age clusters appears contradictory to the 1--8 Gyr signal in the DTD, we note that the intermediate-age clusters in the LMC are generally less massive than old globular clusters, and are therefore less likely to host relatively short-lived objects such as RR Lyrae stars. We quantify this with $n_p$, the expected number of RR Lyrae per cluster given our measured DTD $\Psi T_{vis}$, the mass $M_c$, and age $t_c$ of a cluster:
\begin{equation}
n_p = M_c (\Psi T_{vis})|_{t=t_c}
\label{eq:n_p}
\end{equation}
For each cluster, we measure $\Delta n = (n_o - n_b) - n_p$, i.e. the difference between the observed background-subtracted number of RR Lyrae ($n_o - n_b$) and the number predicted by the DTD. We estimate the uncertainty in $\Delta n$  in the following way: we create 10$^6$ random samples of the 62 clusters, each sample having the same $n_o$ per cluster as given in Table \ref{tab:clusters} but with $n_b$ and $n_p$ generated from a Poisson distribution with mean $n_b$ and $n_p$ values given in Table \ref{tab:clusters}. For any cluster in these samples with a negative value of $n_o - n_b$, we set its $n_o - n_b = 0$. The mean and standard deviation of this random sampling is taken to be the value and error of $\Delta n$. These $\Delta n$ are shown as a function of cluster age and metallicity in Figure \ref{fig:clusters}. We see that $\Delta n$ is consistent with 0 within 2$\sigma$ for most intermediate-age clusters in the LMC. This is because the DTD predicts $\sim 1-3$ RR Lyrae per 10$^{5}$ M$_{\odot}$ of stars, whereas the intermediate-age clusters have an average mass of $4.3 \times 10^4$ M$_{\odot}$, so the majority will have less than 1 RR Lyrae star. This caveat was also raised by \cite{Ols1996} in their study of the SMC: if the RR-Lyrae-rich cluster NGC 121 ($t \approx 12$~Gyr) were scaled down to the magnitude of the 9~Gyr system Lindsey 1 (which has no RR Lyrae), it would host less than 1 RR Lyrae star. The intermediate-age clusters of the Milky Way, including the ones listed above, have the same issue: all have masses less than $10^4$ M$_{\odot}$ \citep{Kalzuny1992, Mermillod2000, Piskunov2008, Grundahl2008, Elsanhoury2016, Kruijssen2019}. Thus, the traditional lower limit on the age of the RR Lyrae progenitor population is subject to small number statistics. 

We note that the globular clusters host a diversity of RR Lyrae populations. NGC 1939, NGC 1916 and NGC 2005 are consistent with $\Delta n = 0$ within their uncertainties, and their production rates $(n_o - n_b)/M_c = (1.8 - 2.5) \times 10^{-5}$ RR Lyrae per M$_{\odot}$ are similar to the measured DTD. The other old clusters, however, host RR Lyrae significantly in excess of their DTD-predicted $n_p$ (i.e. $\Delta n \gg 0$), and have an average production rate $\sim 10^{-4}$ RR Lyrae per M$_{\odot}$, a factor $\sim 5$ times higher than the DTD. 
The high production rate in the old clusters may be an effect of their low metallicities, which is consistent with the differences in production rates of RR Lyrae in the halo versus the bulge and disk of our Galaxy \citep{Layden1995, Dekany2018}, although the large differences in $\Delta n$ for the globular clusters could also be an effect of the second parameter phenomenon in clusters \citep[see][for reviews]{Fusi1997, Catelan09,Dotter2013}. A more detailed assessment of the field and cluster DTDs can be done with a metallicity-dependent DTD that we reserve for future work.

\startlongtable
\begin{deluxetable*}{lccccccccc} \label{tab:clusters}
\tablecaption{LMC Clusters older than 1 Gyr in the \cite{Baumgardt2013} catalog and inside the \cite{Harris2009} region.\label{tab:clusters}}
\tablehead{\colhead{Name} & \colhead{Age (Gyr)} & \colhead{[Fe/H]} & \colhead{$^{a}$Ref} & \colhead{Log Mass (M$_{\odot}$)} & \colhead{$n_o$} & \colhead{$n_b$} & \colhead{$n_p$}}
\startdata
SL569 & 1.2 $\pm$ 0.03 & -0.32 $\pm$ 0.05 & 6 & 4.29 & 0 & 0 & 0 \\
KMK88-38 & 1.26$^{+0.15}_{-0.14}$ & -- & -- & 3.71 & 0 & 0 & 0 \\
BSDL880 & 1.26$^{+0.15}_{-0.14}$ & -- & -- & 3.83 & 0 & 0 & 0 \\
SL197 & 1.26$^{+0.25}_{-0.21}$ & -- & -- & 3.93 & 0 & 0 & 0 \\
HS102 & 1.26$^{+0.15}_{-0.14}$ & -- & -- & 3.87 & 1 & 0 & 0 \\
HS223A & 1.26$^{+0.15}_{-0.14}$ & -- & -- & 4.02 & 0 & 0 & 0 \\
NGC1795 & 1.29$^{+0.16}_{-0.14}$ & -0.23 & 4 & 4.36 & 0 & 0 & 0 \\
NGC1917 & 1.29$^{+0.16}_{-0.14}$ & -0.21 & 4 & 4.42 & 0 & 1 & 1 \\
KMHK898 & 1.32 $\pm$ 0.09 & -- & -- & 4.16 & 0 & 0 & 0 \\
NGC1852 & 1.32$^{+0.16}_{-0.14}$ & -- & -- & 4.51 & 1 & 0 & 0 \\
SL282 & 1.35$^{+0.35}_{-0.28}$ & -- & -- & 3.71 & 1 & 0 & 0 \\
BSDL946 & 1.35$^{+0.2}_{-0.17}$ & -- & -- & 4.17 & 0 & 0 & 0 \\
NGC2154 & 1.41$^{+0.17}_{-0.15}$ & -0.56 & 3 & 4.57 & 1 & 0 & 1 \\
NGC1751 & 1.41$^{+0.17}_{-0.15}$ & -0.44 $\pm$ 0.05 & 6 & 4.6 & 0 & 0 & 1 \\
SL151 & 1.41$^{+0.25}_{-0.21}$ & -- & -- & 4.1 & 0 & 0 & 0 \\
HODGE7 & 1.48$^{+0.18}_{-0.16}$ & -- & -- & 4.47 & 0 & 0 & 0 \\
NGC1846 & 1.48$^{+0.18}_{-0.16}$ & -0.4 & 1 & 5.1 & 1 & 1 & 2 \\
NGC1783 & 1.51$^{+0.18}_{-0.16}$ & -0.75 & 3 & 5.26 & 0 & 1 & 2 \\
NGC1806 & 1.51$^{+0.18}_{-0.16}$ & -0.71 $\pm$ 0.23 & 1 & 5.01 & 1 & 1 & 1 \\
SL136 & 1.55$^{+0.45}_{-0.35}$ & -- & -- & 4.39 & 0 & 0 & 0 \\
NGC2213 & 1.58$^{+0.46}_{-0.35}$ & -0.7 $\pm$ 0.1 & 1 & 4.56 & 0 & 0 & 0 \\
SL180 & 1.58$^{+0.41}_{-0.33}$ & -- & -- & 4.32 & 3 & 0 & 0 \\
H1 & 1.58$^{+0.41}_{-0.33}$ & -0.29 & 5 & 4.84 & 1 & 2 & 1 \\
SL357 & 1.58$^{+0.32}_{-0.27}$ & -- & -- & 4.75 & 0 & 1 & 1 \\
SL390 & 1.58 $\pm$ 0.11 & -0.4 & 2 & 4.48 & 0 & 1 & 0 \\
HS117 & 1.58$^{+0.41}_{-0.33}$ & -- & -- & 4.15 & 2 & 1 & 0 \\
BSDL1102 & 1.62$^{+0.47}_{-0.36}$ & -- & -- & 3.77 & 0 & 0 & 0 \\
SL66 & 1.7$^{+0.7}_{-0.5}$ & -- & -- & 4.34 & 0 & 0 & 0 \\
NGC1652 & 1.7$^{+0.21}_{-0.18}$ & -0.46 & 3 & 4.29 & 0 & 0 & 0 \\
BSDL2652 & 1.74$^{+0.5}_{-0.39}$ & -- & -- & 3.73 & 0 & 0 & 0 \\
HS87 & 1.78$^{+0.22}_{-0.19}$ & -- & -- & 3.98 & 0 & 0 & 0 \\
OGLE-LMC0114 & 1.78$^{+0.46}_{-0.37}$ & -- & -- & 3.71 & 0 & 0 & 0 \\
HS37 & 1.78$^{+1.24}_{-0.73}$ & -- & -- & 4.02 & 1 & 0 & 0 \\
HS177 & 1.78$^{+0.22}_{-0.19}$ & -- & -- & 3.95 & 0 & 0 & 0 \\
H2 & 1.78$^{+0.46}_{-0.37}$ & -0.38 & 5 & 4.98 & 0 & 1 & 1 \\
BSDL734 & 1.78$^{+0.22}_{-0.19}$ & -- & -- & 3.8 & 1 & 0 & 0 \\
KMHK355 & 1.86$^{+0.96}_{-0.63}$ & -- & -- & 3.7 & 0 & 0 & 0 \\
OGLE-LMC0531 & 2.0$^{+0.24}_{-0.22}$ & -- & -- & 3.91 & 0 & 0 & 0 \\
NGC1978 & 2.0$^{+0.24}_{-0.22}$ & -0.38 $\pm$ 0.07 & 1 & 5.33 & 2 & 0 & 3 \\
NGC1651 & 2.0$^{+0.46}_{-0.37}$ & -0.53 $\pm$ 0.03 & 1 & 5.24 & 0 & 0 & 2 \\
SL629 & 2.04$^{+0.1}_{-0.09}$ & -- & -- & 4.0 & 0 & 0 & 0 \\
KMHK1112 & 2.4$^{+0.69}_{-0.54}$ & -- & -- & 3.78 & 1 & 0 & 0 \\
HS88 & 2.45$^{+1.09}_{-0.76}$ & -- & -- & 4.1 & 0 & 0 & 0 \\
SL244 & 2.69 $\pm$ 0.06 & -0.7 $\pm$ 0.2 & 1 & 4.77 & 0 & 0 & 1 \\
SL150 & 2.75 $\pm$ 0.06 & -- & -- & 4.34 & 0 & 0 & 0 \\
HS190 & 2.75$^{+3.41}_{-1.52}$ & -- & -- & 4.51 & 2 & 0 & 1 \\
BSDL2300 & 2.95$^{+0.21}_{-0.2}$ & -- & -- & 4.7 & 0 & 0 & 1 \\
KMHK1188 & 3.02$^{+0.14}_{-0.14}$ & -- & -- & 4.0 & 0 & 0 & 0 \\
H88-93 & 3.02$^{+4.06}_{-1.73}$ & -- & -- & 3.85 & 0 & 0 & 0 \\
BSDL1334 & 3.02$^{+4.74}_{-1.85}$ & -0.4 & 2 & 3.86 & 0 & 0 & 0 \\
SL663 & 3.24$^{+0.48}_{-0.42}$ & -0.7 $\pm$ 0.1 & 1 & 5.23 & 0 & 0 & 2 \\
NGC2121 & 3.24$^{+0.48}_{-0.42}$ & -0.4 $\pm$ 0.1 & 1 & 5.69 & 0 & 1 & 5 \\
NGC2155 & 3.24$^{+0.48}_{-0.42}$ & -0.7 $\pm$ 0.1 & 1 & 4.9 & 0 & 0 & 1 \\
*NGC1939 & 13.49$^{+3.49}_{-2.77}$ & -2.1 $\pm$ 0.19 & 1 & 5.08 & 4 & 1 & 1 \\
*NGC1928 & 13.49$^{+3.49}_{-2.77}$ & -1.27 $\pm$ 0.14 & 1 & 4.87 & 7 & 1 & 1 \\
NGC1898 & 14.13$^{+2.47}_{-2.1}$ & -1.37 $\pm$ 0.15 & 1 & 5.88 & 38 & 2 & 9 \\
NGC1786 & 15.14$^{+0.35}_{-0.34}$ & -2.1 $\pm$ 0.3 & 1 & 5.57 & 47 & 1 & 4 \\
NGC1754 & 15.49$^{+2.29}_{-2.0}$ & -1.42 $\pm$ 0.15 & 1 & 5.39 & 32 & 0 & 3 \\
NGC1916 & 15.85$^{+3.65}_{-2.97}$ & -2.08 & 4 & 5.79 & 14 & 3 & 7 \\
NGC2005 & 16.6$^{+6.31}_{-4.57}$ & -1.35 $\pm$ 0.15 & 1 & 5.49 & 9 & 2 & 4 \\
NGC1835 & 16.6$^{+2.9}_{-2.47}$ & -1.62 $\pm$ 0.15 & 1 & 5.83 & 105 & 2 & 8 \\
NGC2019 & 17.78$^{+3.6}_{-2.99}$ & -1.23 $\pm$ 0.15 & 1 & 5.68 & 49 & 1 & 6
\enddata
\tablecomments{\emph{a} -- References for [Fe/H] measurement. (1) \cite{Harris2009} (2) \cite{Palma2016}, (3) \cite{Girardi2007} (4) \cite{Kontizas1993} (5) \cite{Olszewski1991} (6) \cite{Grocholski2006}\\
* - These 2 clusters are listed as having ages $\sim 1.7$ Gyr in \cite{Baumgardt2013}, but are more likely to be $\gtrsim 10$ Gyr based on HST WFPC2 observations \citep{Mackey2004}.}
\end{deluxetable*}

\section{Implications of the RR Lyrae DTD} \label{sec:implications}
If we assume the DTD measured from the OGLE-IV survey is correct, there are two possible interpretations of this result: 1) RR Lyrae can form from progenitors younger than 10 Gyr \emph{in addition} to the conventional route via older stars, and this result was undetected in previous studies due to various observational limitations, or 2) all OGLE-IV RR Lyrae are older than 10 Gyr stars, and our result is a product of significant (though not readily obvious) systematics in the age derivation of older stellar populations. We discuss both of these interpretations below.

\subsection{Can LMC RR Lyrae have an intermediate-age channel?} \label{sec:canrrlyraebeyoung?}

Since the DTD is generally regarded as a reflection of the progenitor age distribution \citep{Maoz2010, Badenes15}, it is tempting to consider that RR Lyrae in the LMC are being produced by an as-of-yet undiscovered intermediate-age progenitor channel between 1-8 Gyrs, in addition to the usual channel older than 8 Gyrs. 

However, looking at currently available evidence, the possibility of such an undiscovered intermediate-age channel appears to be questionable. Although the small-number statistics described in Section \ref{sec:clusters} is a factor, it is nevertheless true that ancient globular clusters host abundant and rich populations of RR Lyrae compared to the RR Lyrae-poor intermediate-age clusters, a feature easily explained by an exclusively old channel for RR Lyrae formation. Evidence that a small fraction of RR Lyrae may arise from progenitors only a few Gyrs old was recently obtained from thin-disk metal-rich RR Lyrae observed in the \emph{Gaia} data \citep{Zinn2019, Prudil2020, Iorio2020}, and from measurements of companion masses greater than 1 Msun in wide-orbit RR Lyrae binaries \citep{Kervella2019a, Kervella2019b}. The thin disk, metal-rich RR Lyrae population was speculated by \cite{Iorio2020} to be manifestation of binary evolution pulsators \citep{Karczmarek17}, which will register as younger stars. However, the existence of such an intermediate-age channel for LMC RR Lyrae (assuming single-stellar evolution) would be in tension with the age-metallicity relation of the LMC constrained by mutiple studies of field and cluster stars\citep{Cole2005, Carrera2008, Rubele2012, Meschin2014}. According to these studies, the LMC star-formation history between 2-8 Gyrs was associated with [Fe/H] between -0.4 and -1. In contrast, the LMC RR Lyrae are predominantly metal-poor, with [Fe/H]$<-1$ and peaking at [Fe/H]$\sim-1.5$, as confirmed by spectroscopic studies of field RR Lyrae \citep{Gratton2004, Borissova2006}, photometric light curves of RR Lyrae \citep{Haschke2012, Wagner2013, Skowron2016}, and RR Lyrae-hosting globular clusters (Fig \ref{fig:clusters}). Thus the LMC stellar population at ages below 8 Gyrs, where we measure significant signal in the DTD, is more metal-rich than the metallicity range measured for LMC RR Lyrae.

\subsection{Systematic uncertainties in old SADs?}
The DTD represents an empirical connection between the RR Lyrae sample and the SAD map. Since the RR Lyrae sample is highly complete (Section \ref{sec:ogleivsamp}) and has strong independent evidence of originating from old stars (Section \ref{sec:canrrlyraebeyoung?}), and since our DTD recovery method would have correctly recovered a purely old RR Lyrae DTD signal from this SAD map (Section \ref{sec:mockdtd}), it may be possible that the intermediate-age signal in the DTD is indicative of some systematic uncertainty in measuring older stellar ages in the LMC. The source and magnitude of this uncertainty is not readily obvious. The global star-formation history measured by HZ09 is broadly consistent with the interaction history of the Magellanic Clouds derived from proper motion modeling \citep{Lin1995, zaritsky2004b, Besla2007} and with the star formation and chemical enrichment history derived from star clusters \citep{Chilingarian2018}. As shown in Section \ref{sec:sadcompl}, we  verified that any incompleteness or statistical uncertainties in the MCPS photometry is unlikely to be affecting our DTD because: 1) we propagate the reported uncertainties in the SAD map into our DTDs; 2) the DTD retains signal below 8 Gyrs even when measured outside the crowded LMC Bar; 3) wiping out the DTD signal of younger progenitors would require an unreasonably large unseen stellar mass with age $>$8 Gyr; and 4) we directly detect a signal in the DTD below 4 Gyrs, where the main-sequence turnoff is detectable above the MCPS completeness limit.

On the other hand, the SAD solutions per region depend on the overall methodology adopted by the study. An example of this can be found by comparing the \cite{Harris2004} SAD map of the SMC, which was derived using roughly the same methodology used for the LMC, with the \cite{Rubele2018} SAD map of SMC derived from the deeper VISTA near-infrared survey of the Magellanic System \citep[VMC,][]{Kerber2009, Cioni2011}. The \cite{Harris2004} 2--3 Gyr SADs has a distinct ring pattern in the SMC, which they suspected was either the result of photometric incompleteness or systematic uncertainties in the photometric zero-point of the central SMC stars. In the VMC SAD, this ring pattern is absent in the  2--3 Gyr stellar population. Figure 11 of \cite{Rubele2018} shows that the global stellar mass formed at ages $>1$ Gyr in the SMC differ by more than a factor of 2 between \cite{Rubele2018} and \cite{Harris2004}, and the bimodal star-formation history at 2.5 and 10 Gyrs found by \cite{Harris2004} is replaced with a single broad peak at 5 Gyr in \cite{Rubele2018}. These systematic differences could stem from differences in IMF and distance assumed, the stellar isochrone models used, and/or age-metallicity binning. For example, the 
age-metallicity relation measured by the two studies diverges for stars older than 4 Gyrs; this is most likely because \cite{Rubele2018} uses metallicity bins that extend to lower abundances than \cite{Harris2004}. Similar systematic differences in the LMC SADs may also exist. For example, the star-formation history of the Northern Void region measured by \cite{Meschin2014}, using $V$- and $I$-band photometry with the CTIO Blanco 4 m telescope, differs from that of HZ09 for ages younger than 4 Gyr. Partial estimates from upcoming SMASH data reveals a well-mixed SAD in the LMC for ages 

Another source of systematic uncertainty in the SADs may be due to the use of single-stellar evolution isochrone models. According to \cite{Moe2019}, the 
close ($<10$ AU) binary fraction for Milky Way field stars of LMC metallicity is $\sim 30\%$ (compared to $\sim 20\%$ for Solar metallicity), so the influence of binary evolution physics in older populations may be non-negligible. \cite{Stanway2018} have shown that models of integrated spectra and photometry of globular clusters and elliptical galaxies that include binary evolution physics yield age estimates that are different by a few Gyr compared to single-star models. A well-known observational manifestation of binary interaction in old stellar populations is the appearance of blue straggler stars, which are formed from the merger of  $\sim 1 M_{\odot}$ stars.\ Blue straggler stars can mimic younger stars in color-magnitude diagrams as they appear brighter and bluer than the main-sequence turnoff \citep{Santana2016}, and single stellar population models that correct for the presence of blue straggler stars yield globular cluster ages older by a few Gyrs \citep{Fan2012}. Correcting for blue stragglers however is non-trivial as the frequency of blue straggler stars is likely a function of stellar density \citep{Weisz2014, Santana2013, Santana2016}, and the contribution of blue straggler stars at $\sim$Gyr ages have been difficult to determine in composite stellar populations \citep{Surot2019}.

Leaving binary evolution aside, there are uncertainties even in the physics of single-star evolution models that can affect age estimates. For example, \cite{Tayar2017} showed that the mixing length parameter---commonly used to approximate convection theory in 1D stellar models---appears to be correlated with metallicity, and if left unaccounted for when estimating ages from the giant branch (which is the case for ages $>4$ Gyr in the HZ09 maps), it can lead to age uncertainties up to a factor of 2. 

Any number of these reasons could distort the SAD solutions in a subset of the cells, and therefore the final DTD, which is derived from these data.  

\subsection{Caveats and Future work} \label{sec:future}

From our work, we have shown that DTD provides a new rigorous and quantitative test of SADs of Local Group galaxies, in addition to its original purpose as a stellar evolution diagnostic. Although it is possible that unknown sources in systematic uncertainties in the SAD map may be driving our DTD result, we would need further tests to verify the authenticity of this issue, which we will perform in subsequent papers. Estimating the precise form of the HZ09 SAD map that would be consistent with a purely old RR Lyrae DTD is non-trivial because of the large number of SAD parameters involved in this exercise (the stellar masses per age and metallicity bin per cell), and also because RR Lyrae can only reliably constrain the oldest ages. In addition, the coarse age and metallicity binning of the HZ09 map may be less than ideal for constraining the production rate of RR Lyrae if they are coming from a narrower range of ages and metallicities within each bin (as e.g. indicated in Section \ref{sec:clusters}. We can however check if our DTD result persists when calculated with SADs derived from deeper photometric data with finer age and metallicity resolution, such as the upcoming SMASH star-formation histories \citep{RuizLara2020}, as well as in other Local Group galaxies like the SMC \citep{Rubele2018}, M31 \citep{Williams2017} and Local Group dwarfs \citep{Weisz2014}. A more constraining test of the SADs at intermediate and younger ages can also be obtained by calculating DTDs of younger variables stars with well-constrained ages, such as $\delta$-Scutis \citep[ages 1-3 Gyrs,][]{Petersen1996} and Classical Cepheids \citep[ages 70-200 Myrs,][]{Bono2005}. We will pursue these in future papers.

\section{Conclusions}

We have calculated the first delay time-distribution (DTD) of RR Lyrae stars using the large sample of LMC RR Lyrae from the OGLE-IV survey \citep{OGLEIVRRL} and the LMC's SAD map from \cite{Harris2009}. 
Our DTD, shown in Figure \ref{fig:dtd} and Table \ref{tab:rr}, constrains the age-distribution of the full LMC RR Lyrae population, given the measured SAD of the LMC. The
OGLE-IV RR Lyrae sample which overlaps the SAD map of HZ09 contains 29,810 objects, allowing us to recover a DTD with an unprecedented balance of age resolution and detection significance. We determined the DTD signal in each age-bin with an MCMC solver, and used a randomization technique to propagate uncertainties in the SAD map into the final DTD.

Our measured RR Lyrae DTD has statistically significant ($>5\sigma$) power in all age-bins above 1.3 Gyrs, with about 51$\%$ of the RR Lyrae associated with ages between 1.3 and 8 Gyr, and only 46$\%$ with ages above 8 Gyr (the conventional lower limit to RR Lyrae age; note that while the lower limit quoted in the literature is 10 Gyr, the SAD map has a single indivisible age-bin of 8--12 Gyr, and so we refer to the lower limit as 8 Gyr in this paper). 
This would imply that the progenitors of RR Lyrae have zero-age main-sequence masses
$\lesssim 2$ M$_{\odot}$ at LMC metallicity, in contrast with existing constraints. 

We checked the DTD for possible sources of bias. Completeness of the RR Lyrae sample is probably not an issue based on their $I$-band luminosity function and the predominance of RRab (fundamental) pulsators, which are least susceptible to confusion with other types of variables. We also tested our DTD algorithm on fake RR Lyrae maps drawn from a DTD that assumes all RR Lyrae are older than 8 Gyrs, and found that our MCMC algorithm recovers this old DTD without any outlying detections at younger ages. The spatial distribution of RR Lyrae from a purely old DTD is also inconsistent with the spatial distribution of the OGLE-IV RR Lyrae. A possible caveat to our result is the incomplete photometry in the MCPS data and heavy crowding in the LMC central region, which limits the reliability of the SAD maps to ages younger than 4 Gyr; information about older populations are all based on \emph{HST}-derived SADs in a few narrow fields.  However it is not readily obvious how this is producing an intermediate-age signal since: (1) we recover the DTD signal at ages younger than 8 Gyr even after excluding the Bar region, (2) we measure a DTD signal younger than 4 Gyr, and (3) we find that in order to affect the solution, the estimates of the LMC's old stellar mass must be more than an order-of-magnitude greater than current measurements. 

The direct interpretation of our result would be that RR Lyrae have an intermediate-age progenitor channel of 1.3-8 Gyrs stars, in addition to its conventional route via ancient stars, but this possibility is in tension with existing constraints. Both in the Milky Way and Magellanic Clouds, RR Lyrae are abundantly hosted in ancient globular clusters, in contrast with intermediate-age clusters (although this conclusion is somewhat affected by small-number statistics due to the lower-masses of intermediate-age clusters). In addition, the 1.3-8 Gyr RR Lyrae population would likely have [Fe/H]$>$-1 based on existing constraints on the LMC age-metallicity relation, whereas the metallicity distribution of LMC RR Lyrae measured from spectroscopy and photometric light curves is in the range of [Fe/H]$<$-1 with a peak at [Fe/H]$\sim -1.5$. 

The other possibility of the intermediate-age DTD result is the presence of unknown systematics in the LMC SAD map. This is not obvious because the global star-formation history based on the SAD is consistent with the LMC--SMC--Milky Way interaction histories and the chemical enrichment history of the LMC derived from independent studies. However, comparison of the SMC SADs of \cite{Harris2004} and \cite{Rubele2018} shows that spatial solutions of the SAD maps can be influenced by the overall methodology adopted for their  construction (e.g., assumptions about the IMF, the spatial size of cells, the age and metallicity bin size, and stellar isochrone models).\ Age estimation of old stellar populations from color-magnitude diagrams can also be affected by binary evolution processes, such as mass-transfer and mergers, as well as approximations in single-star evolution models. Any combination of these reasons could be skewing the SADs for old stellar populations, and this error could be propagating into the DTD results. We laid out further tests to investigate the physical nature of the systematics that are driving the intermediate-age signal in the DTD, such as revisiting the RR Lyrae DTD once LMC SADs from deeper photometric studies (e.g., SMASH) become available, and also measuring DTDs in SMC and dwarf galaxies, which have less crowded star fields and deep \emph{HST}-derived SADs. We will also continue to apply this technique to other types of variable stars, such as Cepheids and Delta Scutis, to probe other enigmatic phases of stellar evolution.



\section{Acknowledgements}

We are grateful to Horace Smith for reading the manuscript and providing many helpful comments and insight on RR Lyrae observations and models. We also acknowledge the detailed feedback from our anonymous referee on the interpretation of RR Lyrae ages in the LMC and their relation with metallicity, and Dennis Zaritsky, Knut Olsen, Thomas Matheson, Benjamin Williams, J.J.\ Eldridge and Jay Strader for insightful discussions and feedback on this work. 

SKS, CB, and LC are grateful for the suppport of NSF grants AST-1412980 and AST-1907790. CM acknowledges support from the DGAPA/UNAM PAPIIT program grant IG100319 and from the ICC University of Barcelona visiting academic grants and thanks the \emph{Gaia}-UB team for hosting her during part of this research. CM also thanks the Polo de Desarrollo Universitario (PDU) en Ciencias F\'isicas at CURE-UdelaR (Rocha), for their hospitality. DM acknowledges support by grants from the Israel Science Foundation, the German Israeli Science Foundation and the European Research Council (ERC) under the European Union's FP7 Programme, Grant No. 833031.

This work made use of the publicly available OGLE-IV variable star catalog \footnote{http://ogledb.astrouw.edu.pl/~ogle/OCVS/}. This research has made use of NASA's Astrophysics Data System, and the VizieR catalogue access tool, CDS, Strasbourg, France.  The original description of the VizieR service was published in A\&AS 143, 23. This work made use of the IPython package \citep{PER-GRA:2007}, SciPy \citep{scipy2001}, NumPy \citep{numpy}, matplotlib, a Python library for publication quality graphics \citep{Hunter2007}, and Astropy, a community-developed core Python package for Astronomy \citep{Astropy2013}. The Institute for Gravitation and the Cosmos is supported by the Eberly College of Science and the Office of the Senior Vice President for Research at The Pennsylvania State University.

\bibliography{Sarba2018_RRL_arXivSubmission}

\begin{thebibliography}{}
\expandafter\ifx\csname natexlab\endcsname\relax\def\natexlab#1{#1}\fi
\providecommand{\url}[1]{\href{#1}{#1}}
\providecommand{\dodoi}[1]{doi:~\href{http://doi.org/#1}{\nolinkurl{#1}}}
\providecommand{\doeprint}[1]{\href{http://ascl.net/#1}{\nolinkurl{http://ascl.net/#1}}}
\providecommand{\doarXiv}[1]{\href{https://arxiv.org/abs/#1}{\nolinkurl{https://arxiv.org/abs/#1}}}

\bibitem[{{Astropy Collaboration} {et~al.}(2013){Astropy Collaboration},
  {Robitaille}, {Tollerud}, {Greenfield}, {Droettboom}, {Bray}, {Aldcroft},
  {Davis}, {Ginsburg}, {Price-Whelan}, {Kerzendorf}, {Conley}, {Crighton},
  {Barbary}, {Muna}, {Ferguson}, {Grollier}, {Parikh}, {Nair}, {Unther},
  {Deil}, {Woillez}, {Conseil}, {Kramer}, {Turner}, {Singer}, {Fox}, {Weaver},
  {Zabalza}, {Edwards}, {Azalee Bostroem}, {Burke}, {Casey}, {Crawford},
  {Dencheva}, {Ely}, {Jenness}, {Labrie}, {Lim}, {Pierfederici}, {Pontzen},
  {Ptak}, {Refsdal}, {Servillat}, \& {Streicher}}]{Astropy2013}
{Astropy Collaboration}, {Robitaille}, T.~P., {Tollerud}, E.~J., {et~al.} 2013,
  \aap, 558, A33, \dodoi{10.1051/0004-6361/201322068}

\bibitem[{{Badenes} {et~al.}(2015){Badenes}, {Maoz}, \&
  {Ciardullo}}]{Badenes15}
{Badenes}, C., {Maoz}, D., \& {Ciardullo}, R. 2015, Astrophysical Journal
  Letters, 804, L25, \dodoi{10.1088/2041-8205/804/1/L25}

\bibitem[{Badenes {et~al.}(2010)Badenes, Maoz, \& Draine}]{Badenes2010}
Badenes, C., Maoz, D., \& Draine, B.~T. 2010, Monthly Notices of the Royal
  Astronomical Society, 407, 1301, \dodoi{10.1111/j.1365-2966.2010.17023.x}

\bibitem[{{Baumgardt} {et~al.}(2013){Baumgardt}, {Parmentier}, {Anders}, \&
  {Grebel}}]{Baumgardt2013}
{Baumgardt}, H., {Parmentier}, G., {Anders}, P., \& {Grebel}, E.~K. 2013,
  Monthly Notices of the Royal Astronomical Society, 430, 676,
  \dodoi{10.1093/Monthly Notices of the Royal Astronomical Society/sts667}

\bibitem[{{Bertelli} {et~al.}(1994){Bertelli}, {Bressan}, {Chiosi}, {Fagotto},
  \& {Nasi}}]{Bertelli1994}
{Bertelli}, G., {Bressan}, A., {Chiosi}, C., {Fagotto}, F., \& {Nasi}, E. 1994,
  \aaps, 106, 275

\bibitem[{{Besla} {et~al.}(2007){Besla}, {Kallivayalil}, {Hernquist},
  {Robertson}, {Cox}, {van der Marel}, \& {Alcock}}]{Besla2007}
{Besla}, G., {Kallivayalil}, N., {Hernquist}, L., {et~al.} 2007, \apj, 668,
  949, \dodoi{10.1086/521385}

\bibitem[{{Bica} {et~al.}(2008){Bica}, {Bonatto}, {Dutra}, \&
  {Santos}}]{Bica2008}
{Bica}, E., {Bonatto}, C., {Dutra}, C.~M., \& {Santos}, J.~F.~C. 2008, Monthly
  Notices of the Royal Astronomical Society, 389, 678,
  \dodoi{10.1111/j.1365-2966.2008.13612.x}

\bibitem[{{Bildsten} {et~al.}(2012){Bildsten}, {Paxton}, {Moore}, \&
  {Macias}}]{Bildsten2012}
{Bildsten}, L., {Paxton}, B., {Moore}, K., \& {Macias}, P.~J. 2012, \apjl, 744,
  L6, \dodoi{10.1088/2041-8205/744/1/L6}

\bibitem[{{Bitsakis} {et~al.}(2017){Bitsakis}, {Bonfini},
  {Gonz{\'a}lez-L{\'o}pezlira}, {Ram{\'\i}rez-Siordia}, {Bruzual}, {Charlot},
  {Maravelias}, \& {Zaritsky}}]{Bitsakis2017}
{Bitsakis}, T., {Bonfini}, P., {Gonz{\'a}lez-L{\'o}pezlira}, R.~A., {et~al.}
  2017, \apj, 845, 56, \dodoi{10.3847/1538-4357/aa8090}

\bibitem[{{Bono} {et~al.}(2005){Bono}, {Marconi}, {Cassisi}, {Caputo},
  {Gieren}, \& {Pietrzynski}}]{Bono2005}
{Bono}, G., {Marconi}, M., {Cassisi}, S., {et~al.} 2005, \apj, 621, 966,
  \dodoi{10.1086/427744}

\bibitem[{{Borissova} {et~al.}(2006){Borissova}, {Minniti}, {Rejkuba}, \&
  {Alves}}]{Borissova2006}
{Borissova}, J., {Minniti}, D., {Rejkuba}, M., \& {Alves}, D. 2006, \aap, 460,
  459, \dodoi{10.1051/0004-6361:20054132}

\bibitem[{{Bressan} {et~al.}(2012){Bressan}, {Marigo}, {Girardi}, {Salasnich},
  {Dal Cero}, {Rubele}, \& {Nanni}}]{Bressan2012}
{Bressan}, A., {Marigo}, P., {Girardi}, L., {et~al.} 2012, MNRAS, 427, 127,
  \dodoi{10.1111/j.1365-2966.2012.21948.x}

\bibitem[{{Carrera} {et~al.}(2008){Carrera}, {Gallart}, {Hardy}, {Aparicio}, \&
  {Zinn}}]{Carrera2008}
{Carrera}, R., {Gallart}, C., {Hardy}, E., {Aparicio}, A., \& {Zinn}, R. 2008,
  \aj, 135, 836, \dodoi{10.1088/0004-6256/135/3/836}

\bibitem[{{Catelan}(2009)}]{Catelan09}
{Catelan}, M. 2009, Astrophysics and Space Science, 320, 261,
  \dodoi{10.1007/s10509-009-9987-8}

\bibitem[{{Chen} {et~al.}(2014){Chen}, {Girardi}, {Bressan}, {Marigo},
  {Barbieri}, \& {Kong}}]{Chen2014}
{Chen}, Y., {Girardi}, L., {Bressan}, A., {et~al.} 2014, \mnras, 444, 2525,
  \dodoi{10.1093/mnras/stu1605}

\bibitem[{{Chilingarian} \& {Asa'd}(2018)}]{Chilingarian2018}
{Chilingarian}, I.~V., \& {Asa'd}, R. 2018, \apj, 858, 63,
  \dodoi{10.3847/1538-4357/aaba77}

\bibitem[{{Choi} {et~al.}(2016){Choi}, {Dotter}, {Conroy}, {Cantiello},
  {Paxton}, \& {Johnson}}]{Choi2016}
{Choi}, J., {Dotter}, A., {Conroy}, C., {et~al.} 2016, Astrophysical Journal,
  823, 102, \dodoi{10.3847/0004-637X/823/2/102}

\bibitem[{{Cioni} {et~al.}(2011){Cioni}, {Clementini}, {Girardi}, {Guand
  alini}, {Gullieuszik}, {Miszalski}, {Moretti}, {Ripepi}, {Rubele}, {Bagheri},
  {Bekki}, {Cross}, {de Blok}, {de Grijs}, {Emerson}, {Evans}, {Gibson},
  {Gonzales-Solares}, {Groenewegen}, {Irwin}, {Ivanov}, {Lewis}, {Marconi},
  {Marquette}, {Mastropietro}, {Moore}, {Napiwotzki}, {Naylor}, {Oliveira},
  {Read}, {Sutorius}, {van Loon}, {Wilkinson}, \& {Wood}}]{Cioni2011}
{Cioni}, M. R.~L., {Clementini}, G., {Girardi}, L., {et~al.} 2011, \aap, 527,
  A116, \dodoi{10.1051/0004-6361/201016137}

\bibitem[{{Clement} {et~al.}(2001){Clement}, {Muzzin}, {Dufton}, {Ponnampalam},
  {Wang}, {Burford}, {Richardson}, {Rosebery}, {Rowe}, \& {Hogg}}]{Clement2001}
{Clement}, C.~M., {Muzzin}, A., {Dufton}, Q., {et~al.} 2001, Astronomical
  Journal, 122, 2587, \dodoi{10.1086/323719}

\bibitem[{{Cole} {et~al.}(2005){Cole}, {Tolstoy}, {Gallagher}, \&
  {Smecker-Hane}}]{Cole2005}
{Cole}, A.~A., {Tolstoy}, E., {Gallagher}, John~S., I., \& {Smecker-Hane},
  T.~A. 2005, \aj, 129, 1465, \dodoi{10.1086/428007}

\bibitem[{{Conroy}(2013)}]{Conroy2013}
{Conroy}, C. 2013, \araa, 51, 393, \dodoi{10.1146/annurev-astro-082812-141017}

\bibitem[{{Conroy} {et~al.}(2009){Conroy}, {Gunn}, \& {White}}]{Conroy2009}
{Conroy}, C., {Gunn}, J.~E., \& {White}, M. 2009, \apj, 699, 486,
  \dodoi{10.1088/0004-637X/699/1/486}

\bibitem[{{de Grijs} \& {Anders}(2006)}]{deGrijs2006}
{de Grijs}, R., \& {Anders}, P. 2006, \mnras, 366, 295,
  \dodoi{10.1111/j.1365-2966.2005.09856.x}

\bibitem[{{de Marchi} {et~al.}(2007){de Marchi}, {Poretti}, {Montalto},
  {Piotto}, {Desidera}, {Bedin}, {Claudi}, {Arellano Ferro}, {Bruntt}, \&
  {Stetson}}]{ngc6791vars}
{de Marchi}, F., {Poretti}, E., {Montalto}, M., {et~al.} 2007, \aap, 471, 515,
  \dodoi{10.1051/0004-6361:20077386}

\bibitem[{{D{\'e}k{\'a}ny} {et~al.}(2018){D{\'e}k{\'a}ny}, {Hajdu}, {Grebel},
  {Catelan}, {Elorrieta}, {Eyheramendy}, {Majaess}, \&
  {Jord{\'a}n}}]{Dekany2018}
{D{\'e}k{\'a}ny}, I., {Hajdu}, G., {Grebel}, E.~K., {et~al.} 2018, \apj, 857,
  54, \dodoi{10.3847/1538-4357/aab4fa}

\bibitem[{{Demarque} {et~al.}(2004){Demarque}, {Woo}, {Kim}, \&
  {Yi}}]{Demarque2004}
{Demarque}, P., {Woo}, J.-H., {Kim}, Y.-C., \& {Yi}, S.~K. 2004, \apjs, 155,
  667, \dodoi{10.1086/424966}

\bibitem[{{Dotter}(2013)}]{Dotter2013}
{Dotter}, A. 2013, Memorie della Societa Astronomica Italiana, 84, 97.
\newblock \doarXiv{1307.5589}

\bibitem[{{Dotter} {et~al.}(2008){Dotter}, {Chaboyer}, {Jevremovi{\'c}},
  {Kostov}, {Baron}, \& {Ferguson}}]{Dotter2008}
{Dotter}, A., {Chaboyer}, B., {Jevremovi{\'c}}, D., {et~al.} 2008, ApJS, 178,
  89, \dodoi{10.1086/589654}

\bibitem[{{Drake} {et~al.}(2014){Drake}, {Graham}, {Djorgovski}, {Catelan},
  {Mahabal}, {Torrealba}, {Garc{\'{\i}}a-{\'A}lvarez}, {Donalek}, {Prieto},
  {Williams}, {Larson}, {Christen sen}, {Belokurov}, {Koposov}, {Beshore},
  {Boattini}, {Gibbs}, {Hill}, {Kowalski}, {Johnson}, \& {Shelly}}]{Drake2014}
{Drake}, A.~J., {Graham}, M.~J., {Djorgovski}, S.~G., {et~al.} 2014, The
  Astrophysical Journal Supplement, 213, 9, \dodoi{10.1088/0067-0049/213/1/9}

\bibitem[{{Elsanhoury} {et~al.}(2016){Elsanhoury}, {Haroon}, {Chupina},
  {Vereshchagin}, {Sariya}, {Yadav}, \& {Jiang}}]{Elsanhoury2016}
{Elsanhoury}, W.~H., {Haroon}, A.~A., {Chupina}, N.~V., {et~al.} 2016, \na, 49,
  32, \dodoi{10.1016/j.newast.2016.06.002}

\bibitem[{{Fan} \& {de Grijs}(2012)}]{Fan2012}
{Fan}, Z., \& {de Grijs}, R. 2012, \mnras, 424, 2009,
  \dodoi{10.1111/j.1365-2966.2012.21346.x}

\bibitem[{{Foreman-Mackey} {et~al.}(2013){Foreman-Mackey}, {Hogg}, {Lang}, \&
  {Goodman}}]{emcee}
{Foreman-Mackey}, D., {Hogg}, D.~W., {Lang}, D., \& {Goodman}, J. 2013,
  Publications of the Astronomical Society of the Pacific, 125, 306,
  \dodoi{10.1086/670067}

\bibitem[{{Friedmann} \& {Maoz}(2018)}]{Friedmann2018}
{Friedmann}, M., \& {Maoz}, D. 2018, \mnras, 479, 3563,
  \dodoi{10.1093/mnras/sty1664}

\bibitem[{{Fusi Pecci} \& {Bellazzini}(1997)}]{Fusi1997}
{Fusi Pecci}, F., \& {Bellazzini}, M. 1997, in The Third Conference on Faint
  Blue Stars, ed. A.~G.~D. {Philip}, J.~{Liebert}, R.~{Saffer}, \& D.~S.
  {Hayes}, 255

\bibitem[{{Gal-Yam} \& {Maoz}(2004)}]{GalYam2004}
{Gal-Yam}, A., \& {Maoz}, D. 2004, Monthly Notices of the Royal Astronomical
  Society, 347, 942, \dodoi{10.1111/j.1365-2966.2004.07237.x}

\bibitem[{{Gallart} {et~al.}(2005){Gallart}, {Zoccali}, \&
  {Aparicio}}]{Gallart2005}
{Gallart}, C., {Zoccali}, M., \& {Aparicio}, A. 2005, \araa, 43, 387,
  \dodoi{10.1146/annurev.astro.43.072103.150608}

\bibitem[{{Girardi} \& {Marigo}(2007)}]{Girardi2007}
{Girardi}, L., \& {Marigo}, P. 2007, \aap, 462, 237,
  \dodoi{10.1051/0004-6361:20065249}

\bibitem[{{Glatt} {et~al.}(2010){Glatt}, {Grebel}, \& {Koch}}]{Glatt2010}
{Glatt}, K., {Grebel}, E.~K., \& {Koch}, A. 2010, \aap, 517, A50,
  \dodoi{10.1051/0004-6361/201014187}

\bibitem[{{Glatt} {et~al.}(2008){Glatt}, {Gallagher}, {Grebel}, {Nota},
  {Sabbi}, {Sirianni}, {Clementini}, {Tosi}, {Harbeck}, {Koch}, \&
  {Cracraft}}]{Glatt2008}
{Glatt}, K., {Gallagher}, John~S., I., {Grebel}, E.~K., {et~al.} 2008, \aj,
  135, 1106, \dodoi{10.1088/0004-6256/135/4/1106}

\bibitem[{{Gratton} {et~al.}(2004){Gratton}, {Bragaglia}, {Clementini},
  {Carretta}, {Di Fabrizio}, {Maio}, \& {Taribello}}]{Gratton2004}
{Gratton}, R.~G., {Bragaglia}, A., {Clementini}, G., {et~al.} 2004, \aap, 421,
  937, \dodoi{10.1051/0004-6361:20035840}

\bibitem[{{Graur} {et~al.}(2014){Graur}, {Rodney}, {Maoz}, {Riess}, {Jha},
  {Postman}, {Dahlen}, {Holoien}, {McCully}, {Patel}, {Strolger},
  {Ben{\'{\i}}tez}, {Coe}, {Jouvel}, {Medezinski}, {Molino}, {Nonino},
  {Bradley}, {Koekemoer}, {Balestra}, {Cenko}, {Clubb}, {Dickinson},
  {Filippenko}, {Frederiksen}, {Garnavich}, {Hjorth}, {Jones}, {Leibundgut},
  {Matheson}, {Mobasher}, {Rosati}, {Silverman}, {U}, {Jedruszczuk}, {Li},
  {Lin}, {Mirmelstein}, {Neustadt}, {Ovadia}, \& {Rogers}}]{Graur2014}
{Graur}, O., {Rodney}, S.~A., {Maoz}, D., {et~al.} 2014, Astrophysical Journal,
  783, 28, \dodoi{10.1088/0004-637X/783/1/28}

\bibitem[{{Grocholski} {et~al.}(2006){Grocholski}, {Cole}, {Sarajedini},
  {Geisler}, \& {Smith}}]{Grocholski2006}
{Grocholski}, A.~J., {Cole}, A.~A., {Sarajedini}, A., {Geisler}, D., \&
  {Smith}, V.~V. 2006, \aj, 132, 1630, \dodoi{10.1086/507303}

\bibitem[{{Grundahl} {et~al.}(2008){Grundahl}, {Clausen}, {Hardis}, \&
  {Frandsen}}]{Grundahl2008}
{Grundahl}, F., {Clausen}, J.~V., {Hardis}, S., \& {Frandsen}, S. 2008, \aap,
  492, 171, \dodoi{10.1051/0004-6361:200810749}

\bibitem[{{Harris} \& {Zaritsky}(2001)}]{Harris2001}
{Harris}, J., \& {Zaritsky}, D. 2001, \apjs, 136, 25, \dodoi{10.1086/321792}

\bibitem[{{Harris} \& {Zaritsky}(2004)}]{Harris2004}
---. 2004, \aj, 127, 1531, \dodoi{10.1086/381953}

\bibitem[{Harris \& Zaritsky(2009)}]{Harris2009}
Harris, J., \& Zaritsky, D. 2009, Astronomical Journal, 138, 1243,
  \dodoi{10.1088/0004-6256/138/5/1243}

\bibitem[{{Haschke} {et~al.}(2012){Haschke}, {Grebel}, {Duffau}, \&
  {Jin}}]{Haschke2012}
{Haschke}, R., {Grebel}, E.~K., {Duffau}, S., \& {Jin}, S. 2012, \aj, 143, 48,
  \dodoi{10.1088/0004-6256/143/2/48}

\bibitem[{{Hidalgo} {et~al.}(2018){Hidalgo}, {Pietrinferni}, {Cassisi},
  {Salaris}, {Mucciarelli}, {Savino}, {Aparicio}, {Silva Aguirre}, \&
  {Verma}}]{Hidalgo2018}
{Hidalgo}, S.~L., {Pietrinferni}, A., {Cassisi}, S., {et~al.} 2018, \apj, 856,
  125, \dodoi{10.3847/1538-4357/aab158}

\bibitem[{{Holtzman} {et~al.}(1999){Holtzman}, {Gallagher}, {Cole}, {Mould},
  {Grillmair}, {Ballester}, {Burrows}, {Clarke}, {Crisp}, {Evans}, {Griffiths},
  {Hester}, {Hoessel}, {Scowen}, {Stapelfeldt}, {Trauger}, \&
  {Watson}}]{Holtzman1999}
{Holtzman}, J.~A., {Gallagher}, III, J.~S., {Cole}, A.~A., {et~al.} 1999,
  Astronomical Journal, 118, 2262, \dodoi{10.1086/301097}

\bibitem[{{Hunter} {et~al.}(2003){Hunter}, {Elmegreen}, {Dupuy}, \&
  {Mortonson}}]{Hunter2003}
{Hunter}, D.~A., {Elmegreen}, B.~G., {Dupuy}, T.~J., \& {Mortonson}, M. 2003,
  \aj, 126, 1836, \dodoi{10.1086/378056}

\bibitem[{Hunter(2007)}]{Hunter2007}
Hunter, J.~D. 2007, Computing In Science \& Engineering, 9, 90,
  \dodoi{10.1109/MCSE.2007.55}

\bibitem[{{Iorio} \& {Belokurov}(2020)}]{Iorio2020}
{Iorio}, G., \& {Belokurov}, V. 2020, arXiv e-prints, arXiv:2008.02280.
\newblock \doarXiv{2008.02280}

\bibitem[{Jones {et~al.}(2001--)Jones, Oliphant, Peterson,
  {et~al.}}]{scipy2001}
Jones, E., Oliphant, T., Peterson, P., {et~al.} 2001--, {SciPy}: Open source
  scientific tools for {Python}

\bibitem[{{Kaluzny} {et~al.}(2014){Kaluzny}, {Rozyczka}, {Pych}, \&
  {Thompson}}]{ngc6253vars}
{Kaluzny}, J., {Rozyczka}, M., {Pych}, W., \& {Thompson}, I.~B. 2014, \actaa,
  64, 77.
\newblock \doarXiv{1405.5750}

\bibitem[{{Kaluzny} \& {Udalski}(1992)}]{Kalzuny1992}
{Kaluzny}, J., \& {Udalski}, A. 1992, \actaa, 42, 29

\bibitem[{{Karczmarek} {et~al.}(2017){Karczmarek}, {Wiktorowicz},
  {I{\l}kiewicz}, {Smolec}, {St{\c e}pie{\'n}}, {Pietrzy{\'n}ski}, {Gieren}, \&
  {Belczynski}}]{Karczmarek17}
{Karczmarek}, P., {Wiktorowicz}, G., {I{\l}kiewicz}, K., {et~al.} 2017, Monthly
  Notices of the Royal Astronomical Society, 466, 2842, \dodoi{10.1093/Monthly
  Notices of the Royal Astronomical Society/stw3286}

\bibitem[{{Kerber} {et~al.}(2009){Kerber}, {Girardi}, {Rubele}, \&
  {Cioni}}]{Kerber2009}
{Kerber}, L.~O., {Girardi}, L., {Rubele}, S., \& {Cioni}, M.~R. 2009, \aap,
  499, 697, \dodoi{10.1051/0004-6361/200811118}

\bibitem[{{Kervella} {et~al.}(2019{\natexlab{a}}){Kervella}, {Gallenne},
  {Remage Evans}, {Szabados}, {Arenou}, {M{\'e}rand }, {Proto}, {Karczmarek},
  {Nardetto}, {Gieren}, \& {Pietrzynski}}]{Kervella2019a}
{Kervella}, P., {Gallenne}, A., {Remage Evans}, N., {et~al.}
  2019{\natexlab{a}}, \aap, 623, A116, \dodoi{10.1051/0004-6361/201834210}

\bibitem[{{Kervella} {et~al.}(2019{\natexlab{b}}){Kervella}, {Gallenne},
  {Evans}, {Szabados}, {Arenou}, {M{\'e}rand }, {Nardetto}, {Gieren}, \&
  {Pietrzynski}}]{Kervella2019b}
{Kervella}, P., {Gallenne}, A., {Evans}, N.~R., {et~al.} 2019{\natexlab{b}},
  \aap, 623, A117, \dodoi{10.1051/0004-6361/201834211}

\bibitem[{{Kim} {et~al.}(2002){Kim}, {Demarque}, {Yi}, \&
  {Alexander}}]{Kim2002}
{Kim}, Y.-C., {Demarque}, P., {Yi}, S.~K., \& {Alexander}, D.~R. 2002, \apjs,
  143, 499, \dodoi{10.1086/343041}

\bibitem[{{Kinman} \& {Brown}(2010)}]{Kinman2010}
{Kinman}, T.~D., \& {Brown}, W.~R. 2010, Astronomical Journal, 139, 2014,
  \dodoi{10.1088/0004-6256/139/5/2014}

\bibitem[{{Kontizas} {et~al.}(1993){Kontizas}, {Kontizas}, \&
  {Michalitsianos}}]{Kontizas1993}
{Kontizas}, M., {Kontizas}, E., \& {Michalitsianos}, A.~G. 1993, \aap, 269, 107

\bibitem[{{Kruijssen} {et~al.}(2019){Kruijssen}, {Pfeffer}, {Reina-Campos},
  {Crain}, \& {Bastian}}]{Kruijssen2019}
{Kruijssen}, J.~M.~D., {Pfeffer}, J.~L., {Reina-Campos}, M., {Crain}, R.~A., \&
  {Bastian}, N. 2019, \mnras, 486, 3180, \dodoi{10.1093/mnras/sty1609}

\bibitem[{{Layden}(1995)}]{Layden1995}
{Layden}, A.~C. 1995, \aj, 110, 2312, \dodoi{10.1086/117691}

\bibitem[{{Lin} {et~al.}(1995){Lin}, {Jones}, \& {Klemola}}]{Lin1995}
{Lin}, D.~N.~C., {Jones}, B.~F., \& {Klemola}, A.~R. 1995, \apj, 439, 652,
  \dodoi{10.1086/175205}

\bibitem[{{Mackey} \& {Gilmore}(2003)}]{Mackey2003}
{Mackey}, A.~D., \& {Gilmore}, G.~F. 2003, \mnras, 338, 85,
  \dodoi{10.1046/j.1365-8711.2003.06021.x}

\bibitem[{{Mackey} \& {Gilmore}(2004)}]{Mackey2004}
---. 2004, \mnras, 352, 153, \dodoi{10.1111/j.1365-2966.2004.07908.x}

\bibitem[{{Maoz} \& {Badenes}(2010)}]{Maoz2010}
{Maoz}, D., \& {Badenes}, C. 2010, Monthly Notices of the Royal Astronomical
  Society, 407, 1314, \dodoi{10.1111/j.1365-2966.2010.16988.x}

\bibitem[{{Maoz} \& {Graur}(2017)}]{Maoz2017}
{Maoz}, D., \& {Graur}, O. 2017, \apj, 848, 25,
  \dodoi{10.3847/1538-4357/aa8b6e}

\bibitem[{{Maoz} \& {Mannucci}(2012)}]{MaozMan2012}
{Maoz}, D., \& {Mannucci}, F. 2012, Publications of the Astronomical Society of
  Australia, 29, 447, \dodoi{10.1071/AS11052}

\bibitem[{{Maoz} {et~al.}(2012){Maoz}, {Mannucci}, \& {Brandt}}]{Maoz2012b}
{Maoz}, D., {Mannucci}, F., \& {Brandt}, T.~D. 2012, Monthly Notices of the
  Royal Astronomical Society, 426, 3282,
  \dodoi{10.1111/j.1365-2966.2012.21871.x}

\bibitem[{Maoz {et~al.}(2011)Maoz, Mannucci, Li, Filippenko, Valle, \&
  Panagia}]{Maoz2011}
Maoz, D., Mannucci, F., Li, W., {et~al.} 2011, Monthly Notices of the Royal
  Astronomical Society, 412, 1508, \dodoi{10.1111/j.1365-2966.2010.16808.x}

\bibitem[{{Maoz} {et~al.}(2014){Maoz}, {Mannucci}, \& {Nelemans}}]{Maoz2014}
{Maoz}, D., {Mannucci}, F., \& {Nelemans}, G. 2014, Annual Reviews of Astronomy
  and Astrophysics, 52, 107, \dodoi{10.1146/annurev-astro-082812-141031}

\bibitem[{Maoz \& Sharon(2010)}]{Maoz2010a}
Maoz, D., \& Sharon, K. 2010, Astrophysical Journal, 722, 1879,
  \dodoi{10.1088/0004-637X/722/2/1879}

\bibitem[{{Marconi} {et~al.}(2015){Marconi}, {Coppola}, {Bono}, {Braga},
  {Pietrinferni}, {Buonanno}, {Castellani}, {Musella}, {Ripepi}, \&
  {Stellingwerf}}]{Marconi2015}
{Marconi}, M., {Coppola}, G., {Bono}, G., {et~al.} 2015, Astrophysical Journal,
  808, 50, \dodoi{10.1088/0004-637X/808/1/50}

\bibitem[{{Mateu} {et~al.}(2012){Mateu}, {Vivas}, {Downes}, {Brice{\~n}o},
  {Zinn}, \& {Cruz-Diaz}}]{Mateu2012}
{Mateu}, C., {Vivas}, A.~K., {Downes}, J.~J., {et~al.} 2012, Monthly Notices of
  the Royal Astronomical Society, 427, 3374,
  \dodoi{10.1111/j.1365-2966.2012.21968.x}

\bibitem[{Mennekens {et~al.}(2010)Mennekens, Vanbeveren, {De Greve}, \& {De
  Donder}}]{Mennekens2010}
Mennekens, N., Vanbeveren, D., {De Greve}, J.~P., \& {De Donder}, E. 2010,
  Astronomy and Astrophysics, 515, A89, \dodoi{10.1051/0004-6361/201014115}

\bibitem[{{Mermilliod}(2000)}]{Mermillod2000}
{Mermilliod}, J.~C. 2000, Astronomical Society of the Pacific Conference
  Series, Vol. 211, {Massive Clusters in the Milky Way and Magellanic Clouds},
  ed. A.~{Lan{\c{c}}on} \& C.~M. {Boily}, 43

\bibitem[{{Meschin} {et~al.}(2014){Meschin}, {Gallart}, {Aparicio}, {Hidalgo},
  {Monelli}, {Stetson}, \& {Carrera}}]{Meschin2014}
{Meschin}, I., {Gallart}, C., {Aparicio}, A., {et~al.} 2014, \mnras, 438, 1067,
  \dodoi{10.1093/mnras/stt2220}

\bibitem[{{Milone} {et~al.}(2009){Milone}, {Bedin}, {Piotto}, \&
  {Anderson}}]{Milone2009}
{Milone}, A.~P., {Bedin}, L.~R., {Piotto}, G., \& {Anderson}, J. 2009, \aap,
  497, 755, \dodoi{10.1051/0004-6361/200810870}

\bibitem[{{Moe} {et~al.}(2019){Moe}, {Kratter}, \& {Badenes}}]{Moe2019}
{Moe}, M., {Kratter}, K.~M., \& {Badenes}, C. 2019, \apj, 875, 61,
  \dodoi{10.3847/1538-4357/ab0d88}

\bibitem[{{Mosser} {et~al.}(2014){Mosser}, {Benomar}, {Belkacem}, {Goupil},
  {Lagarde}, {Michel}, {Lebreton}, {Stello}, {Vrard}, {Barban}, {Bedding},
  {Deheuvels}, {Chaplin}, {De Ridder}, {Elsworth}, {Montalban}, {Noels},
  {Ouazzani}, {Samadi}, {White}, \& {Kjeldsen}}]{Mosser2014}
{Mosser}, B., {Benomar}, O., {Belkacem}, K., {et~al.} 2014, \aap, 572, L5,
  \dodoi{10.1051/0004-6361/201425039}

\bibitem[{{Nayak} {et~al.}(2016){Nayak}, {Subramaniam}, {Choudhury}, {Indu}, \&
  {Sagar}}]{Nayak2016}
{Nayak}, P.~K., {Subramaniam}, A., {Choudhury}, S., {Indu}, G., \& {Sagar}, R.
  2016, \mnras, 463, 1446, \dodoi{10.1093/mnras/stw2043}

\bibitem[{{Nelemans} {et~al.}(2013){Nelemans}, {Toonen}, \&
  {Bours}}]{Nelemans2013}
{Nelemans}, G., {Toonen}, S., \& {Bours}, M. 2013, in IAU Symposium, Vol. 281,
  Binary Paths to Type Ia Supernovae Explosions, ed. R.~{Di Stefano},
  M.~{Orio}, \& M.~{Moe}, 225--231

\bibitem[{{Olsen}(1999)}]{Olsen1999}
{Olsen}, K.~A.~G. 1999, Astronomical Journal, 117, 2244, \dodoi{10.1086/300854}

\bibitem[{{Olszewski} {et~al.}(1991){Olszewski}, {Schommer}, {Suntzeff}, \&
  {Harris}}]{Olszewski1991}
{Olszewski}, E.~W., {Schommer}, R.~A., {Suntzeff}, N.~B., \& {Harris}, H.~C.
  1991, \aj, 101, 515, \dodoi{10.1086/115701}

\bibitem[{{Olszewski} {et~al.}(1996){Olszewski}, {Suntzeff}, \&
  {Mateo}}]{Ols1996}
{Olszewski}, E.~W., {Suntzeff}, N.~B., \& {Mateo}, M. 1996, Annual Reviews of
  Astronomy and Astrophysics, 34, 511, \dodoi{10.1146/annurev.astro.34.1.511}

\bibitem[{{Palma} {et~al.}(2016){Palma}, {Gramajo}, {Clari{\'a}}, {Lares},
  {Geisler}, \& {Ahumada}}]{Palma2016}
{Palma}, T., {Gramajo}, L.~V., {Clari{\'a}}, J.~J., {et~al.} 2016, \aap, 586,
  A41, \dodoi{10.1051/0004-6361/201527305}

\bibitem[{{Paxton} {et~al.}(2011){Paxton}, {Bildsten}, {Dotter}, {Herwig},
  {Lesaffre}, \& {Timmes}}]{Paxton2011}
{Paxton}, B., {Bildsten}, L., {Dotter}, A., {et~al.} 2011, ApJS, 192, 3,
  \dodoi{10.1088/0067-0049/192/1/3}

\bibitem[{{Paxton} {et~al.}(2013){Paxton}, {Cantiello}, {Arras}, {Bildsten},
  {Brown}, {Dotter}, {Mankovich}, {Montgomery}, {Stello}, {Timmes}, \&
  {Townsend}}]{Paxton2013}
{Paxton}, B., {Cantiello}, M., {Arras}, P., {et~al.} 2013, ApJS, 208, 4,
  \dodoi{10.1088/0067-0049/208/1/4}

\bibitem[{{Paxton} {et~al.}(2015){Paxton}, {Marchant}, {Schwab}, {Bauer},
  {Bildsten}, {Cantiello}, {Dessart}, {Farmer}, {Hu}, {Langer}, {Townsend},
  {Townsley}, \& {Timmes}}]{Paxton2015}
{Paxton}, B., {Marchant}, P., {Schwab}, J., {et~al.} 2015, ApJS, 220, 15,
  \dodoi{10.1088/0067-0049/220/1/15}

\bibitem[{{Paxton} {et~al.}(2018){Paxton}, {Schwab}, {Bauer}, {Bildsten},
  {Blinnikov}, {Duffell}, {Farmer}, {Goldberg}, {Marchant}, {Sorokina},
  {Thoul}, {Townsend}, \& {Timmes}}]{Paxton2018}
{Paxton}, B., {Schwab}, J., {Bauer}, E.~B., {et~al.} 2018, ApJS, 234, 34,
  \dodoi{10.3847/1538-4365/aaa5a8}

\bibitem[{{Pellegrini} {et~al.}(2012){Pellegrini}, {Oey}, {Winkler}, {Points},
  {Smith}, {Jaskot}, \& {Zastrow}}]{Pellegrini2012}
{Pellegrini}, E.~W., {Oey}, M.~S., {Winkler}, P.~F., {et~al.} 2012, \apj, 755,
  40, \dodoi{10.1088/0004-637X/755/1/40}

\bibitem[{P\'erez \& Granger(2007)}]{PER-GRA:2007}
P\'erez, F., \& Granger, B.~E. 2007, Computing in Science and Engineering, 9,
  21, \dodoi{10.1109/MCSE.2007.53}

\bibitem[{{Petersen} \& {Christensen-Dalsgaard}(1996)}]{Petersen1996}
{Petersen}, J.~O., \& {Christensen-Dalsgaard}, J. 1996, \aap, 312, 463

\bibitem[{{Piatti}(2017)}]{Piatti2017}
{Piatti}, A.~E. 2017, \aap, 606, A21, \dodoi{10.1051/0004-6361/201731246}

\bibitem[{{Piatti}(2018)}]{Piatti2018}
---. 2018, \mnras, 475, 2553, \dodoi{10.1093/mnras/stx3344}

\bibitem[{{Pietrinferni} {et~al.}(2004){Pietrinferni}, {Cassisi}, {Salaris}, \&
  {Castelli}}]{Pietrinferni2004}
{Pietrinferni}, A., {Cassisi}, S., {Salaris}, M., \& {Castelli}, F. 2004, \apj,
  612, 168, \dodoi{10.1086/422498}

\bibitem[{{Pietrinferni} {et~al.}(2006){Pietrinferni}, {Cassisi}, {Salaris}, \&
  {Castelli}}]{Pietrinferni2006}
---. 2006, \apj, 642, 797, \dodoi{10.1086/501344}

\bibitem[{{Pietrzynski} \& {Udalski}(2000)}]{Pietrzynski2000}
{Pietrzynski}, G., \& {Udalski}, A. 2000, \actaa, 50, 337.
\newblock \doarXiv{astro-ph/0010360}

\bibitem[{{Piskunov} {et~al.}(2008){Piskunov}, {Schilbach}, {Kharchenko},
  {R{\"o}ser}, \& {Scholz}}]{Piskunov2008}
{Piskunov}, A.~E., {Schilbach}, E., {Kharchenko}, N.~V., {R{\"o}ser}, S., \&
  {Scholz}, R.~D. 2008, \aap, 477, 165, \dodoi{10.1051/0004-6361:20078525}

\bibitem[{{Popescu} {et~al.}(2012){Popescu}, {Hanson}, \&
  {Elmegreen}}]{Popescu2012}
{Popescu}, B., {Hanson}, M.~M., \& {Elmegreen}, B.~G. 2012, \apj, 751, 122,
  \dodoi{10.1088/0004-637X/751/2/122}

\bibitem[{{Pribulla} {et~al.}(2008){Pribulla}, {Rucinski}, {Matthews},
  {Kallinger}, {Kuschnig}, {Rowe}, {Guenther}, {Moffat}, {Sasselov}, {Walker},
  \& {Weiss}}]{m67vars}
{Pribulla}, T., {Rucinski}, S., {Matthews}, J.~M., {et~al.} 2008, \mnras, 391,
  343, \dodoi{10.1111/j.1365-2966.2008.13889.x}

\bibitem[{{Prudil} {et~al.}(2020){Prudil}, {D{\'e}k{\'a}ny}, {Grebel}, \&
  {Kunder}}]{Prudil2020}
{Prudil}, Z., {D{\'e}k{\'a}ny}, I., {Grebel}, E.~K., \& {Kunder}, A. 2020,
  arXiv e-prints, arXiv:2001.02486.
\newblock \doarXiv{2001.02486}

\bibitem[{{Rubele} {et~al.}(2012){Rubele}, {Kerber}, {Girardi}, {Cioni},
  {Marigo}, {Zaggia}, {Bekki}, {de Grijs}, {Emerson}, {Groenewegen},
  {Gullieuszik}, {Ivanov}, {Miszalski}, {Oliveira}, {Tatton}, \& {van
  Loon}}]{Rubele2012}
{Rubele}, S., {Kerber}, L., {Girardi}, L., {et~al.} 2012, \aap, 537, A106,
  \dodoi{10.1051/0004-6361/201117863}

\bibitem[{{Rubele} {et~al.}(2018){Rubele}, {Pastorelli}, {Girardi}, {Cioni},
  {Zaggia}, {Marigo}, {Bekki}, {Bressan}, {Clementini}, {de Grijs}, {Emerson},
  {Groenewegen}, {Ivanov}, {Muraveva}, {Nanni}, {Oliveira}, {Ripepi}, {Sun}, \&
  {van Loon}}]{Rubele2018}
{Rubele}, S., {Pastorelli}, G., {Girardi}, L., {et~al.} 2018, \mnras, 478,
  5017, \dodoi{10.1093/mnras/sty1279}

\bibitem[{{Ruiz-Lara} {et~al.}(2020){Ruiz-Lara}, {Gallart}, {Monelli},
  {Nidever}, {Dorta}, {Choi}, {Olsen}, {Besla}, {Bernard}, {Cassisi},
  {Massana}, {No{\"e}l}, {P{\'e}rez}, {Rusakov}, {Cioni}, {Majewski}, {van der
  Marel}, {Mart{\'\i}nez-Delgado}, {Monachesi}, {Monteagudo}, {Mu{\~n}oz},
  {Stringfellow}, {Surot}, {Vivas}, {Walker}, \& {Zaritsky}}]{RuizLara2020}
{Ruiz-Lara}, T., {Gallart}, C., {Monelli}, M., {et~al.} 2020, \aap, 639, L3,
  \dodoi{10.1051/0004-6361/202038392}

\bibitem[{{Santana} {et~al.}(2013){Santana}, {Mu{\~n}oz}, {Geha},
  {C{\^o}t{\'e}}, {Stetson}, {Simon}, \& {Djorgovski}}]{Santana2013}
{Santana}, F.~A., {Mu{\~n}oz}, R.~R., {Geha}, M., {et~al.} 2013, \apj, 774,
  106, \dodoi{10.1088/0004-637X/774/2/106}

\bibitem[{{Santana} {et~al.}(2016){Santana}, {Mu{\~n}oz}, {de Boer}, {Simon},
  {Geha}, {C{\^o}t{\'e}}, {Guzm{\'a}n}, {Stetson}, \&
  {Djorgovski}}]{Santana2016}
{Santana}, F.~A., {Mu{\~n}oz}, R.~R., {de Boer}, T.~J.~L., {et~al.} 2016, \apj,
  829, 86, \dodoi{10.3847/0004-637X/829/2/86}

\bibitem[{{Schaerer} {et~al.}(1993){Schaerer}, {Meynet}, {Maeder}, \&
  {Schaller}}]{Schaerer1993}
{Schaerer}, D., {Meynet}, G., {Maeder}, A., \& {Schaller}, G. 1993, \aaps, 98,
  523

\bibitem[{{Schaller} {et~al.}(1992){Schaller}, {Schaerer}, {Meynet}, \&
  {Maeder}}]{Schaller1992}
{Schaller}, G., {Schaerer}, D., {Meynet}, G., \& {Maeder}, A. 1992, AAPS, 96,
  269

\bibitem[{{Skowron} {et~al.}(2016){Skowron}, {Soszy{\'n}ski}, {Udalski},
  {Szyma{\'n}ski}, {Pietrukowicz}, {Skowron}, {Poleski}, {Wyrzykowski},
  {Ulaczyk}, {Koz{\l}owski}, {Mr{\'o}z}, \& {Pawlak}}]{Skowron2016}
{Skowron}, D.~M., {Soszy{\'n}ski}, I., {Udalski}, A., {et~al.} 2016, Acta
  Astronomica, 66, 269.
\newblock \doarXiv{1608.00013}

\bibitem[{{Smecker-Hane} {et~al.}(2002){Smecker-Hane}, {Cole}, {Gallagher}, \&
  {Stetson}}]{Smecker2002}
{Smecker-Hane}, T.~A., {Cole}, A.~A., {Gallagher}, III, J.~S., \& {Stetson},
  P.~B. 2002, Astrophysical Journal, 566, 239, \dodoi{10.1086/337985}

\bibitem[{{Smith}(2004)}]{Smith2004}
{Smith}, H.~A. 2004, {RR Lyrae Stars}, 166

\bibitem[{{Soszy{\'n}ski} {et~al.}(2014){Soszy{\'n}ski}, {Udalski},
  {Szyma{\'n}ski}, {Pietrukowicz}, {Mr{\'o}z}, {Skowron}, {Koz{\l}owski},
  {Poleski}, {Skowron}, {Pietrzy{\'n}ski}, {Wyrzykowski}, {Ulaczyk}, \&
  {Kubiak}}]{OGLEIVBULGE}
{Soszy{\'n}ski}, I., {Udalski}, A., {Szyma{\'n}ski}, M.~K., {et~al.} 2014, Acta
  Astronomica, 64, 177.
\newblock \doarXiv{1410.1542}

\bibitem[{{Soszy{\'n}ski} {et~al.}(2016){Soszy{\'n}ski}, {Udalski},
  {Szyma{\'n}ski}, {Wyrzykowski}, {Ulaczyk}, {Poleski}, {Pietrukowicz},
  {Koz{\l}owski}, {Skowron}, {Skowron}, {Mr{\'o}z}, \& {Pawlak}}]{OGLEIVRRL}
---. 2016, Acta Astronomica, 66, 131.
\newblock \doarXiv{1606.02727}

\bibitem[{{Stanway} \& {Eldridge}(2018)}]{Stanway2018}
{Stanway}, E.~R., \& {Eldridge}, J.~J. 2018, \mnras, 479, 75,
  \dodoi{10.1093/mnras/sty1353}

\bibitem[{{Surot} {et~al.}(2019){Surot}, {Valenti}, {Hidalgo}, {Zoccali},
  {S{\"o}kmen}, {Rejkuba}, {Minniti}, {Gonzalez}, {Cassisi}, {Renzini}, \&
  {Weiss}}]{Surot2019}
{Surot}, F., {Valenti}, E., {Hidalgo}, S.~L., {et~al.} 2019, \aap, 623, A168,
  \dodoi{10.1051/0004-6361/201833550}

\bibitem[{{Tayar} {et~al.}(2017){Tayar}, {Somers}, {Pinsonneault}, {Stello},
  {Mints}, {Johnson}, {Zamora}, {Garc{\'\i}a-Hern{\'a}ndez}, {Maraston},
  {Serenelli}, {Allende Prieto}, {Bastien}, {Basu}, {Bird}, {Cohen}, {Cunha},
  {Elsworth}, {Garc{\'\i}a}, {Girardi}, {Hekker}, {Holtzman}, {Huber},
  {Mathur}, {M{\'e}sz{\'a}ros}, {Mosser}, {Shetrone}, {Silva Aguirre},
  {Stassun}, {Stringfellow}, {Zasowski}, \& {Roman-Lopes}}]{Tayar2017}
{Tayar}, J., {Somers}, G., {Pinsonneault}, M.~H., {et~al.} 2017, \apj, 840, 17,
  \dodoi{10.3847/1538-4357/aa6a1e}

\bibitem[{Toonen {et~al.}(2013)Toonen, Nelemans, Bours, Zwart, Claeys,
  Mennekens, \& Ruiter}]{Toonen2013}
Toonen, S., Nelemans, G., Bours, M., {et~al.} 2013, 18th European White Dwarf
  Workshop., 469, 6.
\newblock \doarXiv{1302.0495}

\bibitem[{{Totani} {et~al.}(2008){Totani}, {Morokuma}, {Oda}, {Doi}, \&
  {Yasuda}}]{Totani2008}
{Totani}, T., {Morokuma}, T., {Oda}, T., {Doi}, M., \& {Yasuda}, N. 2008,
  Publications of the Astronomical Society of Japan, 60, 1327, \dodoi{10.1093/
  Publications of the Astronomical Society of Japan/60.6.1327}

\bibitem[{{Udalski} {et~al.}(2015){Udalski}, {Szyma{\'n}ski}, \&
  {Szyma{\'n}ski}}]{Udalski2015}
{Udalski}, A., {Szyma{\'n}ski}, M.~K., \& {Szyma{\'n}ski}, G. 2015, Acta
  Astronomica, 65, 1.
\newblock \doarXiv{1504.05966}

\bibitem[{van~der Walt {et~al.}(2011)van~der Walt, Colbert, \&
  Varoquaux}]{numpy}
van~der Walt, S., Colbert, S.~C., \& Varoquaux, G. 2011, Computing in Science
  \& Engineering, 13, 22, \dodoi{10.1109/MCSE.2011.37}

\bibitem[{{Wagner-Kaiser} \& {Sarajedini}(2013)}]{Wagner2013}
{Wagner-Kaiser}, R., \& {Sarajedini}, A. 2013, \mnras, 431, 1565,
  \dodoi{10.1093/mnras/stt277}

\bibitem[{{Weisz} {et~al.}(2014){Weisz}, {Dolphin}, {Skillman}, {Holtzman},
  {Gilbert}, {Dalcanton}, \& {Williams}}]{Weisz2014}
{Weisz}, D.~R., {Dolphin}, A.~E., {Skillman}, E.~D., {et~al.} 2014, \apj, 789,
  147, \dodoi{10.1088/0004-637X/789/2/147}

\bibitem[{{Williams} {et~al.}(2017){Williams}, {Dolphin}, {Dalcanton}, {Weisz},
  {Bell}, {Lewis}, {Rosenfield}, {Choi}, {Skillman}, \&
  {Monachesi}}]{Williams2017}
{Williams}, B.~F., {Dolphin}, A.~E., {Dalcanton}, J.~J., {et~al.} 2017, \apj,
  846, 145, \dodoi{10.3847/1538-4357/aa862a}

\bibitem[{{Winkler} {et~al.}(2005){Winkler}, {Young}, {Braziunas}, {Condon},
  {Galle}, {Reaser}, {Leiton}, {Smith}, \& {MCELS Team}}]{Winkler2005}
{Winkler}, P.~F., {Young}, A.~L., {Braziunas}, D., {et~al.} 2005, in American
  Astronomical Society Meeting Abstracts, Vol. 207, American Astronomical
  Society Meeting Abstracts, 132.03

\bibitem[{{Yi} {et~al.}(2003){Yi}, {Kim}, \& {Demarque}}]{Yi2003}
{Yi}, S.~K., {Kim}, Y.-C., \& {Demarque}, P. 2003, \apjs, 144, 259,
  \dodoi{10.1086/345101}

\bibitem[{{Zapartas} {et~al.}(2017){Zapartas}, {de Mink}, {Izzard}, {Yoon},
  {Badenes}, {G{\"o}tberg}, {de Koter}, {Neijssel}, {Renzo}, {Schootemeijer},
  \& {Shrotriya}}]{Zapartas17}
{Zapartas}, E., {de Mink}, S.~E., {Izzard}, R.~G., {et~al.} 2017, Astronomy \&
  Astrophysics, 601, A29, \dodoi{10.1051/0004-6361/201629685}

\bibitem[{{Zaritsky} \& {Harris}(2004)}]{zaritsky2004b}
{Zaritsky}, D., \& {Harris}, J. 2004, \apj, 604, 167, \dodoi{10.1086/381795}

\bibitem[{{Zaritsky} {et~al.}(1997){Zaritsky}, {Harris}, \&
  {Thompson}}]{Zaritsky1997}
{Zaritsky}, D., {Harris}, J., \& {Thompson}, I. 1997, \aj, 114, 1002,
  \dodoi{10.1086/118531}

\bibitem[{{Zaritsky} {et~al.}(2004){Zaritsky}, {Harris}, {Thompson}, \&
  {Grebel}}]{zaritsky2004a}
{Zaritsky}, D., {Harris}, J., {Thompson}, I.~B., \& {Grebel}, E.~K. 2004, \aj,
  128, 1606, \dodoi{10.1086/423910}

\bibitem[{{Zhang} {et~al.}(2004){Zhang}, {Deng}, {Zhou}, \& {Xin}}]{ngc188vars}
{Zhang}, X.~B., {Deng}, L., {Zhou}, X., \& {Xin}, Y. 2004, \mnras, 355, 1369,
  \dodoi{10.1111/j.1365-2966.2004.08418.x}

\bibitem[{{Zinn} {et~al.}(2019){Zinn}, {Chen}, {Layden}, \&
  {Casetti-Dinescu}}]{Zinn2019}
{Zinn}, R., {Chen}, X., {Layden}, A.~C., \& {Casetti-Dinescu}, D.~I. 2019,
  \mnras, 3209, \dodoi{10.1093/mnras/stz3580}

\end{thebibliography}

\appendix

\section{Choice of Binning} \label{sec:binning}
Since we are calculating the DTD non-parametrically, the DTD depends on the binning we choose for the ages. The SAD comes with a native resolution of 16 age-bins. We can decide on which combination of these bins provide the optimal DTD through the Bayesian Information Criteria. These are defined as follows,
\begin{equation}
BIC = k\ \mathrm{ln}(N) - 2\ \mathrm{ln}(\mathcal{L}_{max})
\end{equation}
where $k$ is the number of age-bins, $N$ is the number of SAD cells and ln $\mathcal{L}_{max}$ is the maximum likelihood. The binning scheme that minimizes BIC (i.e., the information loss) is favored. We calculate DTDs for different age binnings using just the best-fit SAD (i.e., no randomized SADs) and the original OGLE-IV sample inside the SAD area. 

\begin{figure*}
	\includegraphics[width=\textwidth]{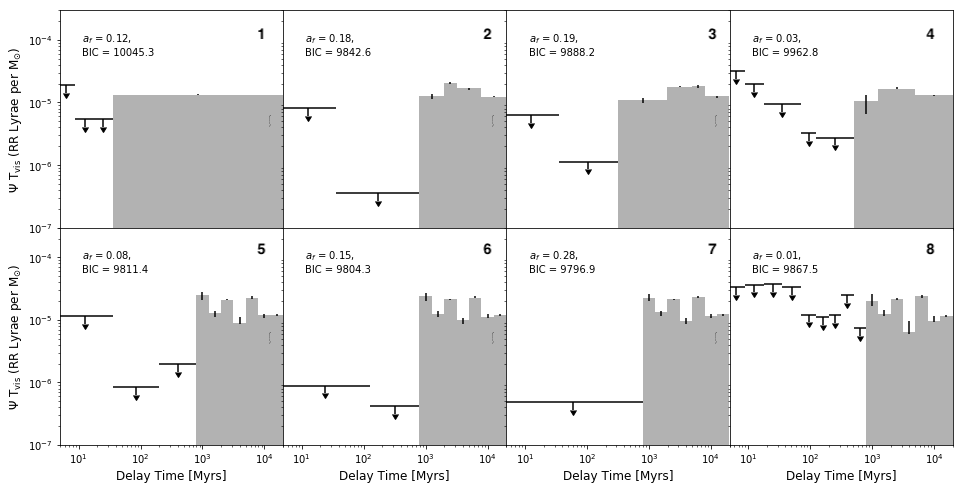}
	\caption{RR Lyrae DTDs calculated with different age binning. The scheme number is shown in the top right corner of each panel. The acceptance fraction $a_f$ of the MCMC solver and the value of the BIC test (BIC) are shown in each panel. Shaded grey regions mark age-bins with significant detections, while arrows show 2$\sigma$ upper limits. We use the binning scheme plotted at bottom right, as it has the smallest information loss, and also has the highest acceptance fraction.}
	\label{fig:bins}
\end{figure*}

We show eight example binning schemes for our DTDs in Figure \ref{fig:bins}, which differ in how the young and old ages are binned. The four binning schemes in the bottom row have the smallest bin-sizes for ages $>0.8$ Gyr and varying bin-sizes for young ages. The schemes in the top row have coarser resolution in the oldest age-bins. More binning schemes similar to these are possible and can be tested, but these eight provide a general sense of the impact of binning. The DTDs measured in the detected bins vary at most by a factor of 2, and are generally around $10^{-5}$ RR Lyrae per M$_{\odot}$. Statistical errors in the DTDs increase with the number of bins, because we are essentially increasing the number of fitting parameters. Binning schemes that retain the highest resolution in the oldest bins measure smaller values of BIC, irrespective of the binning in the younger age-bins. This implies that the RR Lyrae DTD has the strongest signal in the oldest bins, which is consistent with an older stellar origin of RR Lyrae.

Apart from BIC, we also show the acceptance fraction of the MCMC solver, $a_f$ in Figure \ref{fig:bins}. This is the fraction of new steps accepted by the \texttt{emcee} walkers as the scan the multi-dimensional parameter space. While not a model selection statistic like BIC, values of $a_f$ = 0.2--0.5 indicate that the \texttt{emcee} algorithm is performing optimally \citep{emcee}. Values of $a_f$ which are too low indicate multiple peaks in the posterior space separated by valleys of ``low probability", while high values imply that the walkers are simply random-walking, with no regard for the target probability density. We note that while binning schemes in the second row have similar values of BIC, they have different $a_f$, with the native resolution of the SAD (last panel) having the smallest $a_f$. 

Based on these tests, we choose the binning scheme 7 which has the smallest value of BIC, and the largest value of $a_f$ = 0.28 in Figure \ref{fig:bins}.

\section{Uncertainties using the B15 method} \label{app:b15}
\begin{figure}
	\centering
	\includegraphics[width=0.5\textwidth]{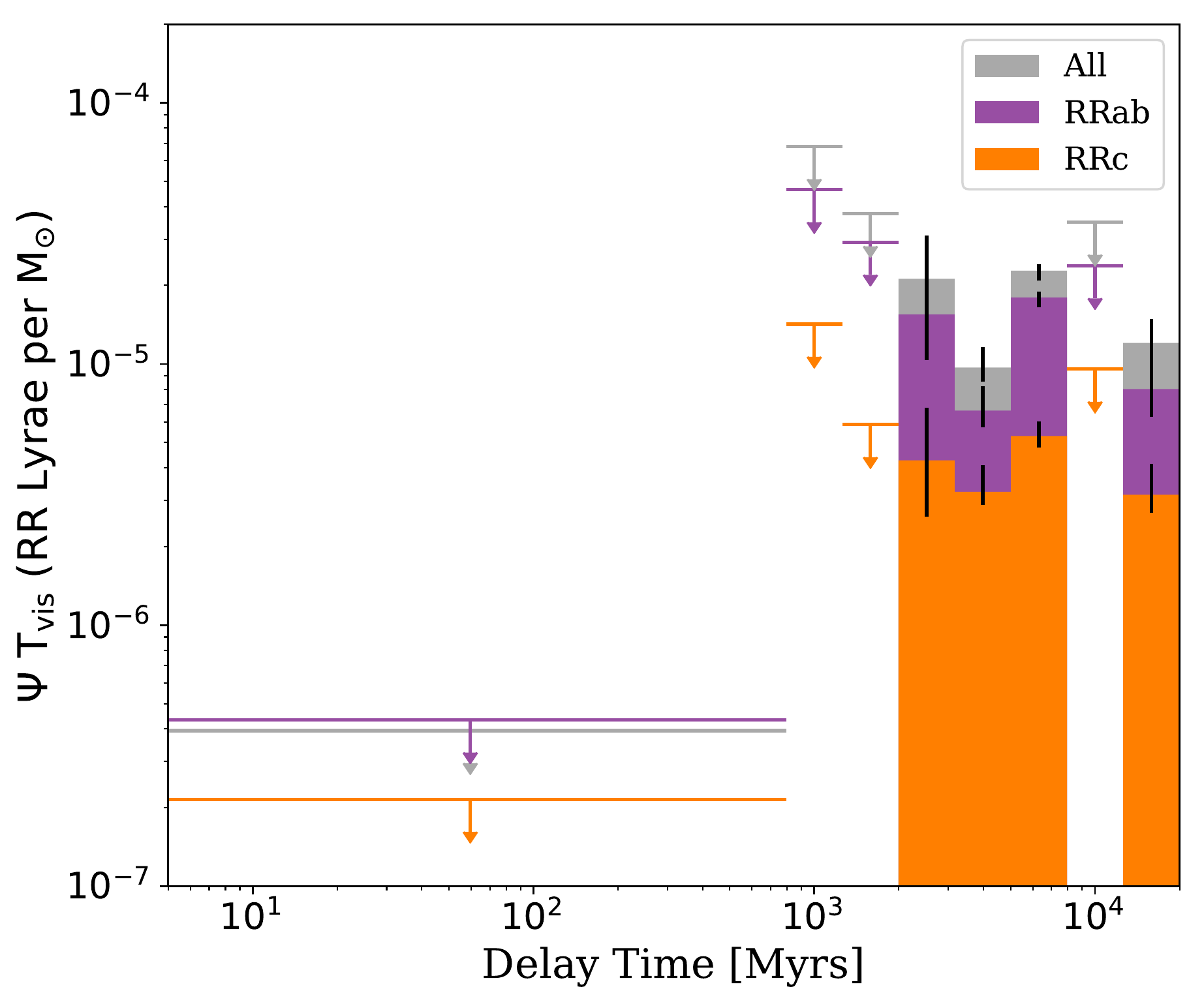}
	\caption{Same as Figure \ref{fig:dtd}, but with uncertainties calculated using the method in \cite{Badenes15}, described in Section \ref{app:b15}.}
	\label{fig:dtdb15}
\end{figure}

When generating the randomized SADs, we assumed that the uncertainties of the stellar masses were normally distributed. However, this is only an approximation, and the underlying shape of the probability distribution of masses is unknown. We therefore also calculated the DTD uncertainties based on the method in B15, which can be treated as a more conservative estimate of the uncertainties. In B15, the 1 $\sigma$ uncertainty due to the SAD is equal to the difference between the DTDs for the best fit SAD, and DTDs for the 68$\%$ upper and lower limits on the SADs. This difference is added in quadrature to the statistical uncertainties in the best-fit DTD, and the total value is used for assessing detectability. The uncertainties using this method are larger, and as a result the DTD (Figure \ref{fig:dtdb15}) is different from the DTD in Figure \ref{fig:dtd}. Nevertheless, it still shows significant signal below 10 Gyrs, particularly in the age range 2-8 Gyrs. The signals in the other age-bins fall below the 2$\sigma$ limit. The total contribution below 10 Gyrs in this case is about 41.6$\%$. Our main science result,$-$ that the DTD has statistically significant signal below 10 Gyrs $-$ is unchanged, even with our most conservative estimates of the uncertainties.
\newpage

\end{document}